\documentclass[aps,prx,reprint,superscriptaddress,amsmath,amssymb,nofootinbib,longbibliography]{revtex4-2}
\usepackage[T1]{fontenc}
\usepackage{lmodern}
\usepackage{mathtools}
\usepackage{amsthm}
\newtheoremstyle{results}{6pt}{6pt}{\normalfont}{}{\bfseries}{.}{.5em}{}
\theoremstyle{results}
\newtheorem{theorem}{Theorem}
\newtheorem{corollary}{Corollary}
\newtheorem{proposition}{Proposition}
\usepackage{booktabs}
\usepackage{bm}
\usepackage{xcolor}
\usepackage{microtype}
\usepackage[colorlinks=true,linkcolor=blue!45!black,citecolor=blue!45!black,urlcolor=blue!45!black]{hyperref}
\usepackage{orcidlink}
\makeatletter
\let\auto@bib@innerbib\@empty
\makeatother
\hypersetup{pdftitle={Probe Geometry and the Cost of Quantum Gate Certification},pdfauthor={Ruge Lin and Guangxi Li},pdfsubject={Quantum gate certification}}
\pdftrailerid{}
\DeclareMathOperator{\Tr}{Tr}
\newcommand{\ket}[1]{\lvert#1\rangle}
\newcommand{\bra}[1]{\langle#1\rvert}
\newcommand{\normF}[1]{\left\lVert#1\right\rVert_F}

\newcommand{\id}{\mathrm{id}}

\newcommand{\HKUSTGZ}{\affiliation{Thrust of Artificial Intelligence, Information Hub, The Hong Kong University of Science and Technology (Guangzhou), Guangzhou 511453, China}}
\newcommand{\QSC}{\affiliation{Quantum Science Center of Guangdong-Hong Kong-Macao Greater Bay Area, Shenzhen 518045, China}}
\allowdisplaybreaks[2]
\begin{document}
\title{Probe Geometry and the Cost of Quantum Gate Certification}
\author{Ruge Lin\,\orcidlink{0000-0002-9297-7622}}
\thanks{Both authors contributed equally to this work.}
\HKUSTGZ
\author{Guangxi Li\,\orcidlink{0009-0009-5481-2411}}
\email[Corresponding author: ]{liguangxi@quantumsc.cn}
\QSC
\begin{abstract}
We relate the geometry of a finite menu of pure probes to the optimal cost of quantum gate certification. For $d<N\le d^2$ distinct probes in dimension $d$, we bound the operator variance gap $\mu_S$ over all sampling weights and mean input states; equality holds exactly for uniform equiangular tight frames. A positive gap gives the sharp global coefficient $1/[(d+1)\mu_S]$ relating mean probe loss to Haar-average gate infidelity. Let $\mu_*$ denote the gap optimized over sampling weights. For a memoryless device and ideal target-output tests, a positive optimized gap gives an optimal worst-case call count asymptotic to $\log(1/\alpha)/[(d+1)\mu_*\delta]$. The infidelity threshold $\delta$ and allowable decision error $\alpha$ approach zero jointly, with dimension and menu fixed. The converse covers adaptive classical choices of fresh probes and collective measurements on retained outputs. For a product-unitary target and exact products of $n$ qubit channels, we exhibit menus of global product probes for which adding a third probe changes the optimal call count from order $n\delta^{-2}\log(1/\alpha)$ to order $\delta^{-1}\log(1/\alpha)$.
\end{abstract}
\maketitle

\section{Introduction}
\label{main-introduction}

Full process tomography reconstructs a channel from tomographically complete data~\cite{poyatos-et-al-1997}. Gate certification can use a smaller set of inputs, but the number of available settings alone does not determine its cost. A menu that identifies the target exactly may respond weakly to small errors and require many repetitions. This paper relates that repetition cost to the geometry of a prescribed finite menu of pure probes.

A verifier tests a memoryless device meant to implement a known unitary gate, using one freshly prepared pure input from the menu per call. The output can be tested by projection onto its ideal ray; the failure probability is the probe's survival loss. Our accuracy criterion is Haar-average gate infidelity, the loss averaged uniformly over all pure inputs. Certification distinguishes the exact target from any fixed implementation whose average infidelity is at least $\delta$, with both decision errors bounded by $\alpha$. We count the worst-case number of device calls.

Consider testing a qubit channel on $\ket0$ and $\ket+=(\ket0+\ket1)/\sqrt2$. If both rays survive exactly, the channel is the identity. A weak dissipative error can still cause survival loss to grow quadratically with its strength while average gate infidelity grows linearly. Testing $\ket1$ reveals the same error at first order. With ideal output comparisons, a risk of order $\delta$ in this family produces a failure signal of order $\delta^2$ with two probes and of order $\delta$ with three.

\paragraph*{The exponent question and earlier verification bounds.}
Mayer and Knill proved that pure probes identify a unitary exactly when they span the system and their overlap graph is connected. For such a set, their general bound on process infidelity is $O(\sqrt{\varepsilon})$ when every probe has survival loss at most $\varepsilon$. They found linear bounds for equiangular tight frames and asked which property of the inputs determines this change of exponent~\cite{mayer-knill-2018}.

Common right-eigenoperators send each probe vector to a multiple of itself; their eigenvalues may differ between rays. Under a common per-probe tolerance, the Trureturing project establishes that an identifying menu has linear small-error robustness exactly when every such operator is scalar, and provides a Lean formalization~\cite{trureturing-2026}. A nonscalar operator supplies a dissipative direction with square-root scaling. At fixed positive sampling weights, the weighted mean loss used here has the same exponents. Theorem~\ref{thm:robustness} and Corollary~\ref{cor:classification} give the sharp weighted form used in our design and certification results.

Zhu and Zhang's formulation of process verification covers arbitrary weighted input ensembles; their spectral bounds apply to balanced ensembles~\cite{zhu-zhang-2020}. For a balanced ensemble with a positive preparation gap, ideal output projectors already give a linear loss-to-risk bound; our variance gap in this case equals their preparation gap divided by $d$. Their framework also identifies projective $2$-designs as optimal verification ensembles. Li et al.\ characterized minimal identifying sets and studied the minimum number of experimental settings~\cite{li-et-al-2021}. These results separate the choice of input settings from the repetitions needed for reliable verification.

\paragraph*{Optimal probe design and certification cost.}
The operator variance gap measures the weakest response of the probes to an error direction. For $d<N\le d^2$ distinct rays, Theorem~\ref{thm:design} bounds this gap over all sampling weights and mean input states. Equality forces uniform weights and an equiangular tight frame. Uniform ETFs attain the bound whenever they exist, establishing Mayer and Knill's fixed-size design conjecture for the optimal linear coefficient under mean probe loss.

For a fixed dimension and menu with a positive optimized gap, Theorem~\ref{thm:certification} identifies the optimal leading call count with ideal target-output projectors as $\delta$ and $\alpha$ jointly tend to zero. Its converse covers adaptive classical choices of fresh probes and collective measurements on retained outputs by constructing one channel that is hard to distinguish from the target on every available input.

\begingroup
\clubpenalty=10000
Theorem~\ref{thm:product} gives a register-wide consequence. For a product-unitary target and exact products of $n$ qubit channels, the two global strings built from $\ket0$ and $\ket+$ require order $n\delta^{-2}\log(1/\alpha)$ calls. Adding $\ket1^{\otimes n}$ reduces this to order $\delta^{-1}\log(1/\alpha)$. Each call applies all $n$ local channels once.
\par
\endgroup

Several earlier constructions provide benchmarks for these bounds. Complementary classical fidelities control process fidelity using two mutually unbiased bases~\cite{hofmann-2005}. Reich et al.\ proposed basis-plus-superposition inputs for gate characterization~\cite{reich-et-al-2013}, and Fiur\'a\v{s}ek and Sedl\'ak obtained a tight finite-error bound for the two-qubit construction~\cite{fiurasek-sedlak-2014}. Baldwin et al.\ studied economical tomography under unitary and near-unitary assumptions~\cite{baldwin-et-al-2014}. Black-box certification of a known unitary has also been studied using trace-norm separation of Choi states~\cite{fawzi-et-al-2023} and diamond-norm separation~\cite{chen-et-al-2026}.

Cao and Bai obtain copy-complexity transitions for Haar-typical pure-state targets under local product measurements~\cite{cao-bai-2026}. Savadatti and Zhang analyze physical fidelity bounds from uncertain Pauli data and optimize shot allocation for fixed certificates~\cite{savadatti-zhang-2026}.

Fidelity objectives are also used in supervised quantum learning~\cite{beer-et-al-2020} and variational quantum compiling~\cite{khatri-et-al-2019}. Caro et al.\ give bounds on statistical generalization from random training data~\cite{caro-et-al-2022}. They also relate locally scrambled input distributions to Haar risk for unitary learning~\cite{caro-et-al-2023}. Our loss-transfer bounds apply to arbitrary completely positive, trace-preserving (CPTP) implementations. For a learned gate, certification uses fresh observations once training has fixed the candidate channel.

\section{The certification task and a two-probe obstruction}
\label{main-model}

\subsection{Losses and physical access}

Let $d\ge2$, and fix a finite weighted ensemble of distinct normalized pure rays,
\begin{equation}
 \begin{aligned}
 S&=\{(p_i,P_i)\}_{i=1}^N,\qquad
 P_i=\ket{\psi_i}\bra{\psi_i},\\
 p_i&>0,\qquad \sum_i p_i=1.
 \end{aligned}
 \label{main-ensemble}
\end{equation}
The target is a unitary $U$ with channel $\mathcal U(X)=UXU^\dagger$. The implementation $\mathcal V$ is an arbitrary CPTP map. The survival test measures the projector $UP_iU^\dagger$ on the device output $\mathcal V(P_i)$. The relative channel $\mathcal E(X)=U^\dagger\mathcal V(X)U$ expresses that output in target coordinates.

All logarithms are natural.

The survival losses and their weighted mean are
\begin{equation}
 \begin{aligned}
 e_i(\mathcal E)&=1-\Tr[P_i\mathcal E(P_i)],\\
 \epsilon_S(\mathcal E)&=\sum_i p_i e_i(\mathcal E).
 \end{aligned}
 \label{main-loss}
\end{equation}
Our test risk is Haar-average gate infidelity,
\begin{equation}
 R(\mathcal E)=\int\left[1-\bra\psi\mathcal E(\ket\psi\bra\psi)\ket\psi\right]d\psi,
 \label{main-risk}
\end{equation}
where $d\psi$ is normalized. For a composite register, the integral ranges over all pure states of the full register, including entangled states. For Kraus matrices $K_a$ of $\mathcal E$, Schumacher's entanglement fidelity at $I/d$~\cite{schumacher-1996} and the average-to-entanglement fidelity relation~\cite{horodecki-et-al-1999,nielsen-2002} give
\begin{equation}
 \begin{aligned}
 F_e(\mathcal E)&=\frac1{d^2}\sum_a|\Tr K_a|^2,\\
 R(\mathcal E)&=\frac d{d+1}[1-F_e(\mathcal E)].
 \end{aligned}
 \label{main-entanglement-fidelity}
\end{equation}
Here $F_e$ is entanglement fidelity relative to the identity. In the normalized Jamio{\l}kowski--Choi representation it is the overlap of the channel state with the maximally entangled state~\cite{jamiolkowski-1972,choi-1975}. Thus process and average infidelities differ by a known dimension factor.

Certification must distinguish $H_0:\mathcal E=\id$, the exact target, from $H_1:R(\mathcal E)\ge\delta$, with each error at most $\alpha\in(0,1/2)$. In the worst case the verifier has at most $T$ calls to the same memoryless channel. On each call, the verifier chooses a classical label $i$ and prepares a fresh $P_i$, which may be entangled within the system. Conditional on $i$, this preparation is independent of the verifier's memory and passes through the device once. On receiving the system output, the verifier may process it jointly with retained outputs and ancillary memory to select later labels or make a collective final measurement.

Conditional on the classical label, the returned output arrives as a tensor factor of the verifier's state before any joint processing. The fidelity proof in Sec.~\ref{main-certification} uses \mbox{this independence.}

Ideal target-output projectors give the sharp call-count coefficients. Section~\ref{main-discussion} gives the precision laws for a verifier supplied with trusted pure target copies.

\subsection{A dissipative error that is hard to detect}

To see why the two probes can miss a first-order error, consider a jump that annihilates $\ket0$ and sends $\ket+$ into its own ray. On $\ket1$, the same jump has a component perpendicular to the input. With equal weights on $S_2=\{\ket0,\ket+\}$, use the following Kraus matrices, with $t$ controlling the strength of the jump:
\begin{equation}
 \begin{aligned}
 L&=\begin{pmatrix}0&1\\0&1\end{pmatrix},\qquad
 K_1(t)=\sqrt t\,L,\\
 K_0(t)&=\begin{pmatrix}1&0\\0&r\end{pmatrix},\qquad
 r=\sqrt{1-2t},\quad 0\le t\le\tfrac12.
 \end{aligned}
 \label{main-qubit-kraus}
\end{equation}
The matrix $K_0(t)$ makes the channel exactly trace-preserving. Write $e_\ell$ for the survival loss on $\ket\ell$. The three survival losses are
\begin{equation}
 e_0=0,\qquad e_+=\frac{(1-r)^2}{4},\qquad e_1=t.
 \label{main-qubit-errors}
\end{equation}
Consequently,
\begin{equation}
 \begin{aligned}
 \epsilon_{S_2}&=\frac{(1-r)^2}{8},\\
 R&=\frac{1-r}{3}+\frac t6
   =\sqrt{2\epsilon_{S_2}}-\frac23\epsilon_{S_2}.
 \end{aligned}
 \label{main-qubit-curve}
\end{equation}
The relation holds exactly throughout the physical interval. Add $\ket1$ and give all three probes equal weights; the resulting mean loss is $\epsilon_{S_3}=[t+(1-r)^2/4]/3$. As $t\to0$, the Haar risk and the two-probe loss behave as $R\sim t/2$ and $\epsilon_{S_2}\sim t^2/8$, respectively. The third probe restores a first-order signal, $\epsilon_{S_3}\sim t/3$.

The channel has a local Stinespring realization~\cite{stinespring-1955} using one initialized environment qubit and a constant number of gates. Apply amplitude damping of probability $2t$, apply a Hadamard when the environment flags a jump, and discard the environment. Appendix~\ref{app-nonunitality-a-local-realization-of-the-qubit-witness} gives the isometry. The system output can still change at first order when the survival loss is quadratic. Section~\ref{main-certification} shows how this small survival signal limits even a verifier who keeps the full outputs.

\section{The linear-versus-square-root transition}
\label{main-gap}

The Kraus identities and channel completion of Ref.~\cite{trureturing-2026} yield the sharp weighted coefficient.

To quantify the cancellation, apply a matrix $A$ to each probe vector and measure the squared length of the component perpendicular to that ray. The weighted mean of these squared lengths is $Q_S(A)$. Its kernel $\mathcal A_S$ is the space of common right-eigenoperators. The operator variance gap $\mu_S$ is the smallest response over traceless matrices of unit Frobenius norm:
\begin{equation}
 \begin{aligned}
 Q_S(A)&=\sum_i p_i\|(I-P_i)A\ket{\psi_i}\|_2^2,\\
 \mathcal A_S&=\{A:A\ket{\psi_i}\in\mathbb C\ket{\psi_i}\ \forall i\},\\
 \mu_S&=\min_{\substack{\Tr A=0\\\normF A=1}}Q_S(A).
 \end{aligned}
 \label{main-gap-definition}
\end{equation}
Here $\|\cdot\|_2$ denotes the Euclidean vector norm, and $\normF A^2=\Tr(A^\dagger A)$. A scalar multiple of $I$ can be added without changing $Q_S(A)$. With positive weights, $\mathcal A_S$ is exactly the kernel of this quadratic form. The minimization allows non-Hermitian matrices, such as the normalized traceless part of the qubit jump $L$. The rays determine which matrices lie in the kernel; reweighting the rays can change the size of a positive gap.

The survival loss sums the squared perpendicular components of the Kraus actions. The Haar risk sums the squared norms of their traceless parts. Explicitly,
\begin{equation}
 \begin{aligned}
 \epsilon_S(\mathcal E)&=\sum_aQ_S(K_a),\\
 R(\mathcal E)&=\frac1{d+1}\sum_a
  \normF{K_a-\frac{\Tr K_a}{d}I}^2.
 \end{aligned}
 \label{main-kraus-identities}
\end{equation}
Trace preservation gives the survival-loss identity. The Haar second moment gives the risk identity (Appendix~\ref{app-robustness-two-exact-kraus-identities}). Applying the gap to the traceless part of each Kraus operator gives the global bound.

\begin{theorem}[Sharp linear robustness]
\label{thm:robustness}
For the fixed pure ensemble with positive weights in Eq.~\eqref{main-ensemble}, a finite coefficient satisfying $R(\mathcal E)\le C\epsilon_S(\mathcal E)$ for every CPTP channel exists if and only if $\mathcal A_S=\mathbb CI$, equivalently $\mu_S>0$. In that case its optimal value is
\begin{equation}
 C_S=\frac1{(d+1)\mu_S}.
 \label{main-optimal-coefficient}
\end{equation}
The inequality holds at every error level, and the optimal ratio is approached by channels tending to the identity.
\end{theorem}

We set $C_S=+\infty$ when $\mu_S=0$. Compactness of the traceless unit sphere gives the equivalence between a positive gap and a scalar kernel.

For necessity and sharpness, embed an arbitrary nonscalar direction in an exact channel:
\begin{equation}
 K_1(t)=\sqrt t\,L,\qquad
 K_0(t)=\sqrt{I-tL^\dagger L},
 \label{main-general-witness}
\end{equation}
valid for $0\le t\le\|L\|_{\mathrm{op}}^{-2}$, where $\|L\|_{\mathrm{op}}$ is the largest singular value. For fixed $L$, Eq.~\eqref{main-kraus-identities} yields
\begin{equation}
 \begin{aligned}
 \epsilon_S(\mathcal E_t)
  &=tQ_S(L)+\frac{t^2}{4}Q_S(L^\dagger L)+O(t^3),\\
 R(\mathcal E_t)
  &=\frac{t}{d+1}\normF{L-\frac{\Tr L}{d}I}^2+O(t^2).
 \end{aligned}
 \label{main-witness-expansion}
\end{equation}
Choosing a normalized traceless gap minimizer gives channels whose risk-to-loss ratio approaches the coefficient in Eq.~\eqref{main-optimal-coefficient}. For a nonscalar kernel direction, the linear term in mean survival loss vanishes while the linear term in risk stays positive. The Kraus form ensures complete positivity, and the square-root completion in $K_0$ ensures trace preservation. The full construction is given~in~\mbox{Appendix~\ref{app-robustness-a-physical-family-that-tests-necessity-and-sharpness}}.

To see why exact identification can coexist with a vanishing gap, join any two rays with nonzero overlap. Zero loss identifies the identity exactly when this graph is connected and the rays span the system~\cite{mayer-knill-2018}. At zero loss each Kraus matrix has every probe as an eigenvector, and trace preservation forces the vectors of Kraus eigenvalues to agree across overlap edges. Connectivity and spanning then make all Kraus matrices scalar. A connected basis can therefore identify the exact channel even though a single jump matrix can have different eigenvalues on the basis rays.

Let $W_S(\varepsilon)=\sup_{\epsilon_S(\mathcal E)\le\varepsilon}R(\mathcal E)$, where the supremum runs over CPTP channels.

\begin{corollary}[Robustness classification]
\label{cor:classification}
For a fixed finite pure ensemble with positive weights, failure to span the system or a disconnected overlap graph gives $W_S(0)>0$. Otherwise, as $\varepsilon\downarrow0$,
\begin{equation}
 W_S(\varepsilon)=
 \begin{cases}
 \Theta_S(\sqrt\varepsilon),&\mu_S=0,\\
 C_S\varepsilon+O_S(\varepsilon^2),&\mu_S>0.
 \end{cases}
 \label{main-trichotomy}
\end{equation}
The constants in $\Theta_S$ and $O_S$ may depend on the fixed ensemble and its dimension.
\end{corollary}

The square-root upper bound is the established connected-spanning bound of Ref.~\cite{mayer-knill-2018}, converted to positive weighted losses using $\max_i e_i\le\epsilon_S/\min_i p_i$. For the matching lower bound in the identifying case, a nonscalar kernel element obeys $Q_S(L^\dagger L)>0$. If this quantity vanished, overlap connectivity would make the Hermitian matrix $L^\dagger L$ scalar, so $L$ would be proportional to a unitary that must also be scalar. Eliminating $t$ in Eq.~\eqref{main-witness-expansion} gives the square-root order. For $\mu_S>0$, solving the linear loss expansion for $t$ gives the leading slope and the quadratic remainder (Appendix~\ref{app-robustness-the-sharp-robustness-trichotomy}).

To decide whether the gap is positive for a spanning menu, choose a basis from the probes and expand the remaining probes in those coordinates. Connect two basis indices whenever an additional probe has nonzero coordinates at both. Such a probe forces the two eigenvalues of a common right-eigenoperator to agree. The gap is positive exactly when this coordinate graph is connected. In particular, a single extra ray with all basis coordinates nonzero restores the linear exponent. Appendix~\ref{app-probe-geometry-a-coordinate-support-criterion} proves the criterion, including the case of nonorthogonal basis rays. Appendix~\ref{app-probe-geometry-a-minimal-identifying-family-in-arbitrary-dimension} gives a square-root witness for the connected basis of Li et al.~\cite{li-et-al-2021} at every fixed dimension.

Direct evaluation of $\mu_S$ from its definition is an eigenvalue problem on the $d^2-1$ dimensional traceless space. The coordinate criterion decides positivity but does not give that value. Optimizing the sampling weights in Eq.~\eqref{main-optimized-gap} is a semidefinite program whose generic matrix constraint acts on the same operator space. Symmetry can reduce this calculation, as in the basis-plus-uniform example below (Appendix~\ref{app-probe-geometry-optimizing-all-sampling-weights}).

\section{Probe design at fixed menu size}
\label{main-design}

When the gap is positive, it gives a survival signal linear in the risk threshold; increasing the gap raises the guaranteed signal at the same threshold. With only $N$ distinct probes, how large a gap can we obtain by choosing their geometry and weights? The mean input $\rho_S=\sum_i p_iP_i$ is unrestricted.

The equality cases involve two standard probe geometries. An equiangular tight frame (ETF) has a maximally mixed uniform mean and constant squared overlap between distinct rays:
\begin{equation}
 \begin{aligned}
 \frac1N\sum_iP_i&=\frac Id,\\
 \Tr(P_iP_j)&=\frac{N-d}{d(N-1)}\quad(i\ne j).
 \end{aligned}
 \label{main-etf}
\end{equation}
A weighted complex projective $2$-design reproduces the second moment of a Haar-random pure input,
\begin{equation}
 \sum_i p_iP_i\otimes P_i=\frac{I+\mathbb F}{d(d+1)},
 \label{main-two-design}
\end{equation}
where $\mathbb F$ swaps the two factors.

\begin{theorem}[Optimal gap at fixed support size]
\label{thm:design}
Let $S$ have $N$ distinct pure rays with positive weights in dimension $d\ge2$. If $N\le d$, then $\mu_S=0$. For $d<N\le d^2$,
\begin{equation}
 \mu_S\le\frac{N-d}{d(N-1)},\qquad
 C_S\ge\frac{d(N-1)}{(d+1)(N-d)}.
 \label{main-support-bound}
\end{equation}
Equality holds exactly for an ETF with uniform weights $1/N$, and is attainable whenever such a frame exists at the specified $(d,N)$. For every support size,
\begin{equation}
 \mu_S\le\frac1{d+1},\qquad C_S\ge1.
 \label{main-absolute-bound}
\end{equation}
Equality in Eq.~\eqref{main-absolute-bound} holds exactly for a weighted complex projective $2$-design and requires $N\ge d^2$.
\end{theorem}

The common squared overlap in Eq.~\eqref{main-etf} saturates the Welch bound~\cite{welch-1974,strohmer-heath-2003}. These frames attain the coefficient of Mayer and Knill's Theorem~3~\cite{mayer-knill-2018} after converting from process infidelity to Haar-average gate infidelity. Theorem~\ref{thm:design} establishes their optimality for the global linear coefficient over arbitrary weights and mean states.

Each probe contributes a rank-one term to the moment matrix $M=\sum_i p_i|P_i\rangle\!\rangle\langle\!\langle P_i|$, where $|P_i\rangle\!\rangle$ is the vectorized projector. Thus $M$ has rank at most $N$. The variance operator $\mathsf Q=\rho_S^{\mathsf T}\otimes I-M$ satisfies $\mathsf Q\succeq\mu_S(I-\Pi_0)$, where $\Pi_0$ projects onto the identity direction. Appendix~\ref{app-optimal-design-operator-representation} derives this representation and the estimate $\mu_S\le\lambda_{\min}(\rho_S)$.

For $0<\mu<\mu_S$, let $H=(\rho_S-\mu I)^{\mathsf T}\otimes I+\mu\Pi_0$. The gap inequality gives $M\preceq H$. The rescaled moment $H^{-1/2}MH^{-1/2}$ has at most $N$ nonzero eigenvalues, all at most one, so its trace is at most $N$. Using the Sherman--Morrison formula~\cite{sherman-morrison-1950}, Appendix~\ref{app-optimal-design-bound-without-a-balance-assumption} bounds this trace from below:
\begin{equation}
 N\ge \Tr(H^{-1}M)\ge\frac{d(1-\mu)}{1-d\mu}.
 \label{main-design-rank-bound}
\end{equation}
The last expression exceeds $d$ for $\mu>0$ and gives Eq.~\eqref{main-support-bound} as $\mu\uparrow\mu_S$. The lower-bound calculation also quantifies how much unequal eigenvalues of $\rho_S$ increase the trace. Retaining that term in the limit forces $\rho_S=I/d$ at equality. The balanced moment then has $N-1$ equal nonzero eigenvalues on the traceless subspace. The corresponding Gram matrix gives the uniform weights and overlaps in Eq.~\eqref{main-etf}. Appendix~\ref{app-optimal-design-equality-and-the-potentially-singular-endpoint} treats the singular endpoint and proves the equality characterization.

The absolute bound follows directly from the trace of $\mathsf Q$. This operator has trace $d-1$ and vanishes on the identity, so its smallest traceless eigenvalue is at most the average of the $d^2-1$ eigenvalues there. Equality makes $\mathsf Q$ scalar on the traceless operator space. Taking its output partial trace first forces balance, and partial transposition then gives Eq.~\eqref{main-two-design}. The associated $M$ has rank $d^2$, recovering the design requirement on the number of rays~\cite{scott-2006}. Zhu and Zhang showed that weighted projective $2$-designs attain the optimum in their process-verification model~\cite{zhu-zhang-2020}. For balanced ensembles and ideal output projectors, our gap is their preparation gap divided by $d$, since $\mathsf Q=I/d-M$ and their preparation operator is $dM$. Appendix~\ref{app-optimal-design-absolute-optimum-and-projective-designs} derives balance and the design condition from equality in the unrestricted bound.

For $N=d+k\le d^2$, the best coefficient is at least $d(d+k-1)/[(d+1)k]$, which is of order $d/k$ when $k\ll d$. A coefficient bounded independently of dimension requires order $d$ extra rays. Two mutually unbiased bases~\cite{wootters-fields-1989} give $C=2d/(d+1)$ using $2d$ rays, matching the complementary-fidelity benchmark~\cite{hofmann-2005,zhu-zhang-2020}. Attaining the absolute optimum $C=1$ requires at least $d^2$ rays.

The uniform $d+1$-ray simplex attains the minimum coefficient $C_S=d^2/(d+1)$ at the smallest support with a positive gap~\cite{mayer-knill-2018}. This construction therefore restores linear robustness with one additional input and a coefficient proportional to dimension. For $d=2^n$, its menu consists of $2^n+1$ product states. Appendix~\ref{app-optimal-design-consequences-and-known-attaining-constructions} gives the construction and the costs of sampling labels and preparing inputs.

Sampling weights can also improve the coefficient on a prescribed menu. Consider the orthonormal basis plus the uniform ray $u=d^{-1/2}\sum_j\ket j$. Averaging over permutations makes the basis weights equal without reducing the gap. This leaves the weight on $u$ as the only variable. Let $\mu_*=\max_p\mu_S$ and $C_*=[(d+1)\mu_*]^{-1}$ for this menu. Weight optimization improves on the uniform coefficient $d+\sqrt{d(d-1)}$, and the optimized coefficient still grows linearly in $d$.

Reich et al.\ introduced the basis-plus-uniform input construction~\cite{reich-et-al-2013}, and Fiur\'a\v{s}ek and Sedl\'ak derived a tight finite-error bound for its two-qubit instance~\cite{fiurasek-sedlak-2014}. Appendix~\ref{app-probe-geometry-spectrum} gives the gap and spectrum, and Appendix~\ref{app-probe-geometry-optimizing-all-sampling-weights} derives the optimum in arbitrary dimension. For the qubit menu $(\ket0,\ket+,\ket1)$, the optimal weights are $(3/10,2/5,3/10)$, giving $\mu_*=1/10$ and $C_*=10/3$.

\section{How a noise promise changes the survival signal}
\label{main-unitality}

A unital relative channel leaves the maximally mixed state unchanged, $\mathcal E(I)=I$. For the two probes in Sec.~\ref{main-model}, this promise restores linear robustness. Throughout this section the finite pure ensemble has positive weights, spans the system, and has a connected overlap graph.

Averaging $Q_S(A)$ and $Q_S(A^\dagger)$ gives the commutator form $G_S(A)=\frac12\sum_i p_i\normF{[A,P_i]}^2$, where $[A,P_i]=AP_i-P_iA$. Minimizing it over traceless matrices of unit Frobenius norm defines the gap $\gamma_S$. Under the stated hypotheses, every matrix that commutes with all probe projectors is scalar, which makes this gap positive (Appendix~\ref{app-nonunitality-the-commutator-form}).

Displacement of the maximally mixed state is an established diagnostic of nonunitality~\cite{garcia-mata-et-al-2005,rastegin-zyczkowski-2014}. We measure it by the trace distance
\begin{equation}
 u(\mathcal E)=\tfrac12\|\mathcal E(I/d)-I/d\|_1.
 \label{main-nonunitality}
\end{equation}
Here $\|X\|_1=\Tr\sqrt{X^\dagger X}$. Given a survival-loss tolerance $\varepsilon$ and a nonunitality budget $\eta$, let $\Phi_S(\varepsilon,\eta)$ be the supremum of $R$ over CPTP channels with $\epsilon_S\le\varepsilon$ and $u\le\eta$.

\begin{proposition}[Sharp robustness under a noise promise]
\label{prop:noise}
For the fixed identifying ensemble above, every unital CPTP channel satisfies
\begin{equation}
 R(\mathcal E)\le\frac{\epsilon_S(\mathcal E)}{(d+1)\gamma_S}.
 \label{main-unital-bound}
\end{equation}
The coefficient is optimal even among unitary alternatives. If $\mu_S=0$, then
\begin{equation}
 \Phi_S(\varepsilon,\eta)
 =\Theta_S\!\left(\varepsilon+\min\{\eta,\sqrt\varepsilon\}\right).
 \label{main-budget-frontier}
\end{equation}
If $\mu_S>0$, then $\Phi_S(\varepsilon,\eta)=\Theta_S(\varepsilon)$. Both frontier estimates hold uniformly for $\eta\ge0$ and sufficiently small $\varepsilon$, with constants depending on the fixed ensemble and dimension.
\end{proposition}

The exact Kraus identity in Eq.~\eqref{app-nonunitality-eq-5} expresses $\sum_aG_S(K_a)$ as the weighted survival loss plus a correction involving $\mathcal E(I)-I$. For a unital channel that correction vanishes, and the commutator gap gives Eq.~\eqref{main-unital-bound}. A Hermitian gap minimizer generates unitary channels whose risk-to-loss ratio approaches the coefficient~\mbox{(Appendix~\ref{app-nonunitality-sharpness-within-the-unital-class})}.

For a general CPTP channel, $\mathcal E(I)-I$ has zero trace. Its trace pairing with $\rho_S$ is therefore controlled by the spectral range $\Delta_\rho=\lambda_{\max}(\rho_S)-\lambda_{\min}(\rho_S)$, yielding
\begin{equation}
 R(\mathcal E)\le
 \frac{\epsilon_S(\mathcal E)+(d/2)\Delta_\rho u(\mathcal E)}
 {(d+1)\gamma_S}.
 \label{main-budget-bound}
\end{equation}
For balanced ensembles, $\rho_S=I/d$, the correction vanishes and $\mu_S=\gamma_S$. Every identifying ensemble with $\mu_S=0$ is therefore unbalanced.

The frontier's upper bound combines Eq.~\eqref{main-budget-bound} with square-root robustness. A depolarizing channel supplies the linear lower term. The nonscalar kernel witness in Eq.~\eqref{main-general-witness} has quadratic survival loss and nonunitality linear in $t$. Choosing $t$ proportional to $\min\{\eta,\sqrt\varepsilon\}$ gives the second term. For a positive gap, Theorem~\ref{thm:robustness} and the depolarizing example give the two linear bounds~\mbox{(Appendix~\ref{app-nonunitality-the-sharp-joint-error-frontier})}.

For the two-probe example, $\gamma_{S_2}=1/4$ gives the sharp unital bound $R\le4\epsilon_{S_2}/3$, and the dissipative witness has $u(\mathcal E_t)=t/\sqrt2$.

For an identifying menu with zero gap, the frontier is linear under exact unitality or a budget of order $\varepsilon$. A budget between $\varepsilon$ and $\sqrt\varepsilon$ sets the larger term; if the budget is fixed and positive, the behavior is square-root as $\varepsilon\to0$. The upper bound in Eq.~\eqref{main-budget-bound} scales with $1/\gamma_S$. To check the budget directly, prepare the maximally mixed state $I/d$ and bound the trace distance between the output and $I/d$.

\section{Optimal call counts for a fixed menu}
\label{main-certification}

Fix the allowed rays $P_1,\ldots,P_N$. Define $T_{\mathrm{opt}}(\delta,\alpha)$ to be the smallest worst-case call cap for the certification task of Sec.~\ref{main-model}, with both decision errors bounded by $\alpha$. A fresh probe is selected classically for each call and sent once through the same memoryless channel. Under the model of Sec.~\ref{main-model}, the verifier may select labels adaptively, retain arbitrary quantum memory, and measure \mbox{outputs collectively.}

The mean survival loss depends on how often each ray is sampled. Write $Q_i(A)=\|(I-P_i)A\ket{\psi_i}\|_2^2$ and let $\mathsf Q_i$ represent this form on the traceless Hilbert--Schmidt operator space. The sampling distribution $p$ belongs to the simplex $\Delta_N=\{p\in\mathbb R^N:p_i\ge0,\sum_i p_i=1\}$, which permits some rays to receive zero weight. The optimized gap and its loss coefficient are
\begin{equation}
 \mu_* =\max_{p\in\Delta_N}\lambda_{\min}
 \left(\left.\sum_i p_i\mathsf Q_i\right|_{\Tr A=0}\right),
 \qquad c=(d+1)\mu_*.
 \label{main-optimized-gap}
\end{equation}
\begin{theorem}[Optimal leading certification cost]
\label{thm:certification}
Suppose $\mu_*>0$ and ideal target-output projectors are available. For the permitted verifiers, at fixed dimension $d$ and fixed finite menu,
\begin{equation}
 \lim_{\substack{\delta\to0\\\alpha\to0}}
 \frac{\delta T_{\mathrm{opt}}(\delta,\alpha)}{\log(1/\alpha)}
 =\frac1{(d+1)\mu_*}.
 \label{main-certification-coefficient}
\end{equation}
The infidelity threshold $\delta$ and allowable decision error $\alpha$ may tend to zero independently.
\end{theorem}

The upper bound follows from independent sampling with immediate survival tests. For the converse, we construct one alternative channel that is hard to distinguish on every available input. Appendix~\ref{app-certification} contains the full argument and the finite-sample bounds.

\subsection{Sampling weights and a common hard channel}

Strictly positive weights can approach the same optimum by mixing with a positive distribution. If $\mu_*>0$, then $0<c\le1$ and an optimizing distribution gives $\epsilon_S\ge cR$.

Draw labels from the optimizing distribution and use the independent all-pass rule of state and process verification~\cite{pallister-et-al-2018,zhu-zhang-2020}. Test each device output against its target ray, accepting only if every test passes. The ideal channel always passes; for a bad channel, the false-acceptance probability is at most $(1-c\delta)^T$. Thus a sufficient call cap for the all-pass protocol is
\begin{equation}
 T_{\mathrm{pass}}=\left\lceil\frac{\log(1/\alpha)}{-\log(1-c\delta)}\right\rceil.
 \label{main-cert-upper}
\end{equation}

Let $Z$ be a density operator on the complex traceless matrix space and take both traces below over that space. The smallest eigenvalue of the weighted operator is the minimum of its expectation over such $Z$. Finite-dimensional convex minimax~\cite{sion-1958} then gives
\begin{equation}
 \mu_* =\min_{\substack{Z\succeq0\\\Tr Z=1}}
          \max_i\Tr(\mathsf Q_iZ),
 \label{main-minimax}
\end{equation}
Factor an optimizer as $Z=\sum_a|L_a\rangle\!\rangle\langle\!\langle L_a|$, where the vectors represent matrices with $\Tr L_a=0$ and $\sum_a\normF{L_a}^2=1$. This mixture of error directions satisfies $\sum_aQ_i(L_a)\le\mu_*$ on every input (Appendix~\ref{app-certification-fixed-menu-minimax-duality}).

This minimizing mixture defines the adversarial \mbox{channel through}
\begin{equation}
 \begin{aligned}
 B&=\sum_aL_a^\dagger L_a,\qquad
 K_a(t)=\sqrt t\,L_a,\\
 K_0(t)&=\sqrt{I-tB},\qquad 0\le t\le1.
 \end{aligned}
 \label{main-multi-kraus}
\end{equation}
Since $B\succeq0$ and $\Tr B=1$, the square root is defined throughout the interval and the Kraus operators give an exact CPTP channel $\mathcal E_t$. The identities for Haar risk and probe loss give the simultaneous bounds
\begin{equation}
 R(\mathcal E_t)\ge\frac{t}{d+1},\qquad
 \max_i e_i(\mathcal E_t)\le t\mu_*+t^2.
 \label{main-adversary-bound}
\end{equation}
The traceless Kraus terms give the risk lower bound and the linear term in the probe losses. The remaining loss satisfies $Q_i(K_0)\le\|K_0-I\|_F^2\le t^2$ (Appendix~\ref{app-certification-exact-multi-kraus-adversary}).

\subsection{The fidelity bound for adaptive protocols}

Use squared fidelity $F(\rho,\sigma)=\|\sqrt\rho\sqrt\sigma\|_1^2$~\cite{uhlmann-1976,jozsa-1994}. The corresponding root fidelity is $f=\sqrt F$, with $f(X,Y)=\|\sqrt X\sqrt Y\|_1$ also defined for subnormalized positive blocks~\cite{watrous-2018}. Let $\rho_i,\sigma_i$ be the outputs under the two channel hypotheses for input $i$, with $F(\rho_i,\sigma_i)\ge1-b$ and $0\le b<1$. Denote the unconditional final verifier states by $\Omega_0,\Omega_1$. Under the fresh-input model, any protocol with at most $T$ calls satisfies
\begin{equation}
 F(\Omega_0,\Omega_1)\ge(1-b)^T.
 \label{main-adaptive-fidelity}
\end{equation}
This bound specializes the classical--quantum amortization identity of Wilde et al.\ (Lemma~26) to sandwiched R\'enyi order $1/2$~\cite{wilde-et-al-2020}. At each round, a common instrument selects a classical label and produces subnormalized memory blocks $X_i,Y_i$ that include the possibly different label probabilities under the two hypotheses. The input is fresh conditional on the label, so the corresponding output appends as a tensor factor. Block additivity and tensor multiplicativity then give the new root fidelity $f_{\mathrm{new}}$
\begin{equation}
 \begin{aligned}
 f_{\mathrm{new}}&=\sum_i f(X_i,Y_i)f(\rho_i,\sigma_i)\\
 &\ge\sqrt{1-b}\sum_i f(X_i,Y_i).
 \end{aligned}
 \label{main-block-fidelity}
\end{equation}
Common quantum processing cannot decrease fidelity. We therefore iterate the bound from identical initial states, padding any branch that has stopped with further calls and discarding their outputs. Squaring the resulting bound gives Eq.~\eqref{main-adaptive-fidelity}; Appendix~\ref{app-certification-adaptive-cq-fidelity-lemma} gives the proof with arbitrary retained memory.

The outcome distributions of a binary test with both errors at most $\alpha$ have fidelity at most $4\alpha(1-\alpha)$. For $b>0$, measurement monotonicity therefore gives
\begin{equation}
 T\ge\frac{\log[1/(4\alpha(1-\alpha))]}{-\log(1-b)}.
 \label{main-fidelity-call-bound}
\end{equation}
If the verifier must always accept the target, the numerator improves to $\log(1/\alpha)$. For $b=0$ the two devices are indistinguishable under the allowed inputs.

Set $t=(d+1)\delta$ in Eq.~\eqref{main-adversary-bound}. The resulting channel has risk at least $\delta$ and squared fidelity with the pure ideal output of at least $1-b_\delta$ on every input, where $b_\delta=c\delta+(d+1)^2\delta^2$. For small $\delta$, Eq.~\eqref{main-fidelity-call-bound} with $b=b_\delta$ and the all-pass cap bracket the optimal count (Appendix~\ref{app-certification-finite-sample-sandwich}). Salek et al.\ established asymptotic error exponents for fixed pairs of classical--quantum channels~\cite{salek-et-al-2022}. Adaptivity can improve the Stein exponent for composite classical channel hypotheses~\cite{bergh-et-al-2023}. Here the minimax step chooses one alternative from the composite risk class whose per-input bound holds across the menu.

Both logarithmic denominators in these finite bounds are $c\delta+O_{d,S}(\delta^2)$, and the ratio of their logarithmic numerators tends to one. This proves Theorem~\ref{thm:certification}. Appendix~\ref{app-certification-exact-optimized-sampling-coefficient} gives the fixed-confidence bounds and the exact coefficient when the verifier must always accept the target.

The support theorem also makes Eq.~\eqref{main-certification-coefficient} a menu-size lower bound for adaptive verifiers. For a fixed identifying menu the optimal orders are $\Theta(\delta^{-1}\log(1/\alpha))$ when the kernel is scalar and $\Theta(\delta^{-2}\log(1/\alpha))$ otherwise, uniformly for $0<\alpha\le\alpha_*<1/2$ at fixed $\alpha_*$. In the zero-gap case, Eq.~\eqref{main-general-witness} gives the quadratic upper bound on every probe loss needed for the converse. Square-root robustness gives the matching all-pass upper bound. For a nonidentifying menu a positive-risk unitary can give exactly the same outputs as the target, making certification below that risk impossible (Appendix~\ref{app-certification-complete-precision-trichotomy}).

\section{Two versus three global probes for independent qubits}
\label{main-product}

We return to the three rays of the qubit example to see how their choice changes the cost of certifying an entire register. Let $D=2^n$ and suppose the relative \mbox{channel is exactly}
\begin{equation}
 \mathcal E=\bigotimes_{j=1}^n\mathcal E_j,
 \label{main-product-promise}
\end{equation}
where the target is a product unitary. The qubit channels $\mathcal E_j$ are fixed across calls and may differ between sites. One global call sends a fresh $n$-qubit product input through all $n$ local channels once and returns all $n$ outputs. Consider the two menus $S_2=\{\ket0^{\otimes n},\ket+^{\otimes n}\}$ and $S_3=S_2\cup\{\ket1^{\otimes n}\}$.

Let $T_k^\star(\delta,\alpha)$ be the smallest worst-case number of these whole-register calls for the menu $S_k$, with both decision errors at most $\alpha$ under the fresh-input access model of Sec.~\ref{main-model}. Let $Y_r=e_{r^{\otimes n}}(\mathcal E)$ be the survival loss on the global input $\ket r^{\otimes n}$, for $r=0,+,1$. For the ordered menu $(0,+,1)$, set
\[
 \begin{aligned}
 p&=(3/10,2/5,3/10),\\
 \epsilon_p&=\sum_r p_rY_r,\qquad C_n=\frac{5D}{D+1}.
 \end{aligned}
\]
The risk $R$ remains the full-register Haar-average infidelity, including entangled test states.

\begin{theorem}[Certification of independent qubits]
\label{thm:product}
Under the exact-product promise in Eq.~\eqref{main-product-promise}, with ideal target-output projectors available, the following statements hold.

\textnormal{(i)} For every fixed $0<\alpha_*<1/2$, constants independent of $n,\delta,\alpha$ give
\begin{equation}
 \begin{aligned}
 T_2^\star&=\Theta\!\left(n\delta^{-2}\log(1/\alpha)\right),\\
 T_3^\star&=\Theta\!\left(\delta^{-1}\log(1/\alpha)\right)
 \end{aligned}
 \label{main-product-sample-laws}
\end{equation}
for all $n\ge1$, $0<\delta\le1/8$, and $0<\alpha\le\alpha_*$. The two-direction lower bound also holds when all $2^n$ tensor strings of local $\ket0,\ket+$ inputs are available.

\textnormal{(ii)} Every channel satisfying the product promise obeys
\begin{equation}
 R(\mathcal E)\le C_n\epsilon_p(\mathcal E).
 \label{main-product-coefficient}
\end{equation}
The coefficient $C_n$ is the smallest global linear coefficient over all sampling weights on $S_3$. For each fixed $n$,
\begin{equation}
 \lim_{\substack{\delta\to0\\\alpha\to0}}
 \frac{\delta T_3^\star(\delta,\alpha)}{\log(1/\alpha)}=C_n,
 \label{main-product-leading}
\end{equation}
where $\delta$ and $\alpha$ may tend to zero independently.
\end{theorem}

The proof uses exact product identities and the local qubit bounds. For $P_r=\ket r\bra r$, write $x_j=1-F_e(\mathcal E_j)$ and $y_{jr}=e_r(\mathcal E_j)$, with vectors $x=(x_1,\ldots,x_n)$ and $y_r=(y_{1r},\ldots,y_{nr})$. Define $F(z)=1-\prod_j(1-z_j)$ for $z\in[0,1]^n$. Then
\begin{equation}
 R=\frac D{D+1}F(x),\qquad Y_r=F(y_r).
 \label{main-product-identities}
\end{equation}
If both global inputs in $S_2$ survive perfectly, every local channel preserves both $\ket0$ and $\ket+$. Those two rays identify the local target, so $S_2$ identifies the product target exactly. Appendix~\ref{app-product-channels} proves the quantitative bounds.

For the two-probe upper bound, set $\epsilon_2=(Y_0+Y_+)/2$. By local square-root robustness, each $x_j$ is at most a constant times $\sqrt{y_{j0}+y_{j+}}$. Combining Cauchy--Schwarz with the product identities gives $R\le C\sqrt{n\epsilon_2}$ with an absolute constant (Appendix~\ref{app-product-channels-upper-bound}). Risk at least $\delta$ therefore implies mean survival loss at least $\delta^2/(C^2n)$. The all-pass protocol detects this loss using the two \mbox{uniform strings.}

For the two-probe lower bound, use the qubit witness in Eq.~\eqref{main-qubit-kraus} at every site. With $s=\sqrt{1-2t}$, the common local entanglement infidelity is $x(t)=(1-s)/2+t/4$. The local survival losses on $\ket0$ and $\ket+$ are $y_0=0$ and $y_+=(1-s)^2/4$. Choose $t$ by
\begin{equation}
 x(t)=1-\left(1-\frac{D+1}{D}\delta\right)^{1/n},
 \label{main-product-calibration}
\end{equation}
so that the global risk is exactly $\delta$. This calibration gives $t$ of order $\delta/n$ uniformly in the stated range. The local plus loss is quadratic in $t$, and even the all-plus \mbox{input has loss}
\begin{equation}
 b=1-(1-y_+)^n=O(\delta^2/n).
 \label{main-product-hardness}
\end{equation}
Every mixed string of the two local inputs has survival loss at most $b$, because its local losses are either $0$ or $y_+$. Equation~\eqref{main-fidelity-call-bound} then gives the matching lower order for all those strings (Appendix~\ref{app-product-channels-lower-bound-and-its-uniform-calibration}).

For the three-probe bound, the optimal qubit weights give
$x_j\le(3/2)y_{j0}+2y_{j+}+(3/2)y_{j1}$. The \mbox{product inequality}
\begin{equation}
 \begin{gathered}
 x_j\le\sum_r c_ry_{jr},\quad c_r\ge1\\
 \Longrightarrow\quad F(x)\le\sum_r c_rF(y_r)
 \end{gathered}
 \label{main-product-inequality}
\end{equation}
transfers this estimate to the register (Appendix~\ref{app-product-channels-an-exact-three-probe-inequality}). Since the local coefficients are $5p_r$, it gives Eq.~\eqref{main-product-coefficient} at every error level.
With noise confined to one site, the mean probe loss is unchanged and the local risk-to-loss ratio can approach its sharp value. Converting local Haar risk to register risk gives the corresponding global coefficient. Sampling the three global strings assigns the same weights to the three local rays; the locally optimal weights therefore also optimize $C_n$ on this global menu. The all-pass protocol gives the upper bound on the three-probe call count. For the converse, a depolarizing alternative at one site has risk $\delta$ and loss of order $\delta$ on every probe. It gives the matching lower order uniformly in $n$ (Appendix~\ref{app-product-channels-three-direction-sample-complexity-and-exact-coefficient}).

For the exact coefficient, put the optimized qubit minimax adversary at one site and leave the other sites noiseless. If $R_{\rm local}$ is the risk at that site, the full-register risk is $R=g_D R_{\rm local}$ with $g_D=3D/[2(D+1)]$. For small $\delta$, setting $t=3\delta/g_D$ in the qubit construction with $\mu_*=1/10$ gives full-register risk at least $\delta$ and loss at most $\delta/C_n+O(\delta^2)$ on every allowed probe. The lower and upper bounds therefore have the same leading term, proving Eq.~\eqref{main-product-leading} at fixed $n$ (Appendix~\ref{app-product-channels-three-direction-sample-complexity-and-exact-coefficient}).

The finite bounds give the call counts in Table~\ref{main-table-certification}. Here the necessary count for two directions is Eq.~\eqref{main-fidelity-call-bound} rounded up. Evaluate $b$ using the exact calibration in Eq.~\eqref{main-product-calibration} and the exact all-plus loss in Eq.~\eqref{main-product-hardness}. For three probes, the sufficient count is the all-pass cap in Eq.~\eqref{main-cert-upper} with $c=1/C_n$ and weights $(3/10,2/5,3/10)$.

\begin{table}[htb]
 \centering
 \setlength{\tabcolsep}{4pt}
 \begin{tabular}{rrrr}
 \toprule
 Qubits $n$ & Risk $\delta$ & \shortstack[r]{Two directions\\Calls required at least} & \shortstack[r]{Three probes\\Calls sufficient} \\
 \midrule
 8 & 5\% & 11,312 & 297 \\
 \textbf{32} & \textbf{5\%} & \textbf{45,489} & \textbf{299} \\
 128 & 5\% & 181,831 & 299 \\
 \addlinespace
 32 & 10\% & 10,792 & 149 \\
 32 & 2.5\% & 186,626 & 598 \\
 \bottomrule
 \end{tabular}
 \caption{Finite call bounds for exact products of $n$ qubit channels, with a product-unitary target, ideal target-output projectors and decision error $\alpha=0.05$. The two-direction menu uses $\ket0^{\otimes n}$ and $\ket+^{\otimes n}$. The three-probe menu adds $\ket1^{\otimes n}$. Each whole-register call uses all $n$ local channels.}
 \label{main-table-certification}
\end{table}

At $\delta=0.05$, increasing $n$ raises the necessary count for two directions while the sufficient count for three probes stays near 300. At $n=32$, decreasing $\delta$ raises both bounds, with a larger increase for the two-direction \mbox{lower bound.}

At fixed small global accuracy and confidence, the change is from order $n$ global repetitions to order one. Total local channel uses are respectively $\Theta(n^2\delta^{-2}\log(1/\alpha))$ and $\Theta(n\delta^{-1}\log(1/\alpha))$. The all-pass protocols use product preparations and local survival measurements with $O(n)$ ideal primitives per global call.

The product identities rely on exact tensor factorization. Experimental crosstalk measurements can probe departures from independent operation~\cite{sarovar-et-al-2020}.
For arbitrary CPTP channels, the $0$-fidelity $F_0$ averages survival fidelities over $4^n$ product rays built from single-qubit symmetric informationally complete (SIC) measurements~\cite{greenaway-et-al-2021}, and Mayer proved the dimension-independent bound $1-F_e\le(3/2)(1-F_0)$~\cite{mayer-2021}. For product channels, the three-probe construction gives dimension-independent precision bounds with matching lower bounds.

\section{Statistical precision and resources beyond gate calls}
\label{main-discussion}

Threshold-based gate verification can allow failures under the good hypothesis and has been demonstrated with local operations~\cite{zhang-et-al-2022}. Let $q=\epsilon_S$ be the rejection probability of a randomized survival test. Suppose $q\le\theta m$ for a good channel and $q\ge m$ for a bad channel, with fixed $0\le\theta<1$ and $0<m\le1$. Reject when the observed failure frequency exceeds $(1+\theta)m/2$. Both errors are at most $\alpha$ after
\begin{equation}
 \left\lceil\frac{8\log(1/\alpha)}{(1-\theta)^2m}\right\rceil
 \label{main-shrinking-tolerance}
\end{equation}
independent trials (Appendix~\ref{app-certification-shrinking-acceptance-tolerance}). The loss bounds give $m$ of order $\delta$ in the linear regime and $\delta^2$ in the singular identifying regime. Allowing the good channel's survival loss to shrink on this scale preserves the respective precision exponents.

The situation changes around a fixed noisy baseline. Take depolarizing channels whose Haar infidelities are $r_0$ and $r_0+\delta$, keeping $0<r_0<(d-1)/d$ fixed and taking $\delta>0$ sufficiently small. At infidelity $r$, a pure input $P$ is mapped to $(1-r)P+r(I-P)/(d-1)$. The two outputs commute; their squared fidelity has a quadratic deficit whose coefficient, $1/[4r_0(1-r_0)]$, is one quarter of the classical Fisher information at $r_0$~\cite{braunstein-caves-1994}. On any menu, the adaptive converse gives
\begin{equation}
 T=\Omega\!\left(r_0(1-r_0)\delta^{-2}\log(1/\alpha)\right)
 \label{main-noisy-baseline}
\end{equation}
for distinguishing this pair, uniformly for $0<\alpha\le\alpha_*<1/2$ at fixed $\alpha_*$ (Appendix~\ref{app-certification-fixed-noisy-baselines-have-a-different-exponent}).

With one fresh trusted pure target copy per call, a SWAP comparison of $\sigma_i=\mathcal V(P_i)$ with $\tau_i=UP_iU^\dagger$ gives the antisymmetric outcome with probability $(1-\Tr\sigma_i\tau_i)/2=e_i/2$~\cite{barnett-et-al-2003}. Rejecting whenever this outcome occurs preserves both precision exponents. Twice the leading call bound of the ideal-projector protocol suffices (Appendix~\ref{app-certification-trusted-pure-target-references}).

Direct fidelity estimation samples observables~\cite{flammia-liu-2011,da-silva-et-al-2011}, and randomized benchmarking infers error from sequence decay~\cite{magesan-et-al-2011}. With a dimension assumption, repeated gate sequences can certify a qubit phase gate with inverse-infidelity sample complexity~\cite{noller-et-al-2025}. Miklin et al.\ use repeated sequences to certify a qubit $\sqrt{X}$ gate up to a unitary change of basis, with uncharacterized preparation and readout in a time-independent qubit model~\cite{miklin-et-al-2026}. The present call bounds concern one pass through the fixed device per fresh probe.

With ideal input preparations and target-output projectors, Theorem~\ref{thm:certification} gives the general leading-coefficient limit at fixed dimension and fixed finite menu with a positive optimized gap. Each probe is prepared independently of the verifier's retained memory conditional on its classical label, then passes once through the same memoryless channel. Theorem~\ref{thm:product} additionally assumes a product-unitary target and exact tensor factorization of the relative channel across qubits. That promise makes both product-probe survival probabilities and entanglement fidelities factor across sites, yielding call orders uniform in the register size. The exact leading coefficient in this product model is stated at fixed $n$.

Further questions concern the exact finite-error envelope $W_S(\varepsilon)$ for general menus and the optimal certification costs away from the small-error limit. The global linear inequality and finite call bounds constrain these quantities. Their full shapes can depend on the ensemble. Other directions include taking joint limits in dimension and precision, comparing with imperfect reference states, and deriving bounds on the effects of departures from exact product structure or independence between calls.

\section{Conclusion}
\label{main-conclusion}

The operator variance gap links finite-probe geometry to the cost of quantum gate certification. A positive gap gives the sharp global coefficient relating weighted mean survival loss to Haar-average gate infidelity, accounting for both the probe rays and their sampling weights.

The support bound makes the tradeoff between probe settings and repetitions explicit. For $d<N\le d^2$, it holds for arbitrary weights and mean input states, and equality characterizes uniformly weighted equiangular tight frames. Whenever such a frame exists, it minimizes the global linear coefficient at that support size. In particular, $d+1$ probes suffice for linear robustness, but their best coefficient is $d^2/(d+1)$; keeping the coefficient bounded as $d$ grows requires $N-d$ to grow at least proportionally to $d$.

For a fixed dimension and menu with positive optimized gap, independent sampling and ideal target-output tests attain the optimal leading call count as the infidelity threshold and decision error tend jointly to zero. The matching converse includes adaptive classical probe choices and collective measurements on retained outputs within the stated fresh-probe model.

For a product-unitary target and exact products of $n$ qubit channels, adding $\ket1^{\otimes n}$ to the two global probes $\ket0^{\otimes n}$ and $\ket+^{\otimes n}$ changes the optimal call count from order $n\delta^{-2}\log(1/\alpha)$ to order $\delta^{-1}\log(1/\alpha)$, uniformly in $n$. Product preparations and local survival measurements realize both protocols with $O(n)$ ideal primitives per whole-register call. The change in repetition cost comes from improving the response to error directions that the two-probe menu detects only weakly.

\begingroup
\makeatletter
\def\section{\@startsection{section}{1}{\z@}
  {.5cm \@plus .2ex \@minus .1ex}{.2cm}
  {\normalfont\small\bfseries\centering}}
\makeatother
\begin{acknowledgments}
This work was partially supported by the National Natural Science Foundation of China (Grant No.~62402206), the Guangdong Provincial Quantum Science Strategic Initiative (Grant No.~GDZX2503001), and the Shenzhen Science and Technology Program (Grant No.~ZDYJ20251211120859002).
\end{acknowledgments}

\section*{Use of artificial intelligence}
Several models in OpenAI's GPT-6 and GPT-6.2 families were used throughout the development of this work. The authors set the scientific direction and scope, guided the investigation, and selected and refined the results and arguments presented in the manuscript. Iterative exchanges with these models contributed to identifying and formulating the research problem, developing preliminary theorem statements, and generating proof strategies and proof drafts. The models also assisted with deriving explicit examples, searching and comparing the literature, developing and running numerical checks, and drafting and revising the manuscript. The authors made the final scientific and editorial decisions and take full responsibility for the correctness, originality, and presentation \mbox{of the work.}

\section*{Data availability}
The analytical notes, numerical checks, and supporting data are available in the project repository~\cite{quantum-training-robustness-repository}.

\par
\endgroup
\clearpage
\onecolumngrid
\appendix

\section{Operator geometry and sharp robustness}
\label{app-robustness}

Fix the dimension $d\geq 2$ and the finite ensemble $S=\{(p_i,P_i)\}_{i=1}^N$. The weights satisfy $p_i>0$ and $\sum_i p_i=1$. For a unitary target $U$, the relative CPTP channel $\mathcal E(X)=U^\dagger\mathcal V(X)U$ expresses the system output in target coordinates.

We compare probe loss and Haar risk through two Kraus identities, then construct exact channels that show when the resulting bound is sharp. Appendix~\ref{app-probe-geometry} turns the kernel condition into a test on the probe rays.

\subsection{Two exact Kraus identities}
\label{app-robustness-two-exact-kraus-identities}

Write $P_i=|\psi_i\rangle\langle\psi_i|$, with normalized $\psi_i$, and define
\begin{equation}
\label{app-robustness-eq-1}
e_i(\mathcal E)=1-\Tr[P_i\mathcal E(P_i)],
\qquad
\epsilon_S(\mathcal E)=\sum_{i=1}^N p_i e_i(\mathcal E),
\qquad
R(\mathcal E)=1-F_{\mathrm{avg}}(\mathcal E,\id).
\end{equation}

The probe variance measures how much of $A\psi_i$ lies outside each probe ray. The decomposition $A\psi_i=P_iA\psi_i+(I-P_i)A\psi_i$ is orthogonal, giving
\begin{equation}
\label{app-robustness-eq-2}
\begin{aligned}
Q_S(A)
&=\sum_i p_i\left(\|A\psi_i\|^2-
      |\langle\psi_i|A|\psi_i\rangle|^2\right)\\
&=\sum_i p_i\|(I-P_i)A\psi_i\|^2.
\end{aligned}
\end{equation}

This is a positive semidefinite quadratic form on the full complex matrix space, with the Hilbert--Schmidt inner product. Since $(I-P_i)I\psi_i=0$,
\begin{equation}
\label{app-robustness-eq-3}
Q_S(A+zI)=Q_S(A),\qquad z\in\mathbb C.
\end{equation}

For any Kraus representation $\{K_a\}$~\cite{kraus-1971}, projecting the output onto $P_i$ gives pass probability $\sum_a|\langle\psi_i|K_a|\psi_i\rangle|^2$. Trace preservation, $\sum_aK_a^\dagger K_a=I$, makes $\sum_a\|K_a\psi_i\|^2$ equal to one. Subtracting the pass probability gives the loss,
\begin{equation}
\label{app-robustness-eq-4}
e_i(\mathcal E)
=\sum_a\left(\|K_a\psi_i\|^2-
 |\langle\psi_i|K_a|\psi_i\rangle|^2\right).
\end{equation}

Taking the weighted mean gives
\begin{equation}
\label{app-robustness-eq-5}
\boxed{\epsilon_S(\mathcal E)=\sum_a Q_S(K_a).}
\end{equation}

For a Haar-random ray $P=\ket\psi\bra\psi$, write $|\bra\psi A\ket\psi|^2=\Tr[(A\otimes A^\dagger)(P\otimes P)]$. Each $P\otimes P$ projects onto the symmetric vector $\psi\otimes\psi$, and the Haar average is invariant under every $V\otimes V$ with $V$ unitary. This action is irreducible on the symmetric subspace, so the average is the projector $(I+\mathbb F)/2$ divided by the dimension $d(d+1)/2$. The identity contributes $|\Tr A|^2$ to the trace and the swap $\mathbb F$ contributes $\Tr(AA^\dagger)=\|A\|_F^2$. Thus
\begin{equation}
\label{app-robustness-eq-6}
\int |\langle\psi|A|\psi\rangle|^2\,d\psi
=\frac{\|A\|_F^2+|\Tr\,A|^2}{d(d+1)}.
\end{equation}

Here $\|\cdot\|_F$ is the Frobenius norm. Subtract the sum of the averaged pass probabilities from one and use $\sum_a\|K_a\|_F^2=d$ from trace preservation. This gives $(d-d^{-1}\sum_a|\Tr K_a|^2)/(d+1)$. Removing the scalar projection $(\Tr K_a)I/d$ subtracts $|\Tr K_a|^2/d$ from each squared norm, so
\begin{equation}
\label{app-robustness-eq-7}
\boxed{
R(\mathcal E)
=\frac{1}{d+1}\sum_a
\left\|K_a-\frac{\Tr\,K_a}{d}I\right\|_F^2.
}
\end{equation}

Equivalently, if $F_e$ is entanglement fidelity with respect to the identity,
\begin{equation}
\label{app-robustness-eq-8}
F_e(\mathcal E)=\frac{1}{d^2}\sum_a|\Tr\,K_a|^2,
\qquad
R(\mathcal E)=\frac{d}{d+1}[1-F_e(\mathcal E)].
\end{equation}

\subsection{Proof of Theorem~\ref{thm:robustness}}
\label{app-robustness-the-linear-robustness-theorem}

Define the common right-eigenoperator space and its gap by
\begin{equation}
\label{app-robustness-eq-9}
\mathcal A_S
=\{A:A\psi_i\in\mathbb C\psi_i\ \text{for every }i\},
\qquad
\mu_S
=\min_{\substack{\Tr\,A=0\\\|A\|_F=1}}Q_S(A).
\end{equation}

Every squared term has positive weight, so $Q_S(A)=0$ exactly when all perpendicular components vanish, giving $\ker Q_S=\mathcal A_S$. Subtracting the scalar part and normalizing a nonscalar matrix in this kernel gives a traceless matrix of unit Frobenius norm and zero variance. Conversely, a zero gap supplies such a matrix by compactness of the unit sphere. Hence
\begin{equation}
\label{app-robustness-eq-10}
\mu_S>0\quad\Longleftrightarrow\quad\mathcal A_S=\mathbb CI.
\end{equation}

We prove Theorem~\ref{thm:robustness} by deriving the global bound and then constructing physical channels that test necessity and sharpness. When $\mu_S>0$, the candidate coefficient is
\begin{equation}
\label{app-robustness-eq-11}
C_S=\frac{1}{(d+1)\mu_S}.
\end{equation}

\subsubsection{Proof of the upper bound}
\label{app-robustness-proof-of-the-upper-bound}

Set $K_a^0=K_a-\Tr(K_a)I/d$. The loss is invariant under this shift. Applying the gap to each normalized traceless part and scaling back gives $Q_S(K_a^0)\geq\mu_S\|K_a^0\|_F^2$, also valid when $K_a^0=0$. Summing gives the global bound,
\begin{equation}
\label{app-robustness-eq-12}
\epsilon_S(\mathcal E)
=\sum_a Q_S(K_a^0)
\geq\mu_S\sum_a\|K_a^0\|_F^2
=(d+1)\mu_S R(\mathcal E).
\end{equation}

\subsubsection{A physical family that tests necessity and sharpness}
\label{app-robustness-a-physical-family-that-tests-necessity-and-sharpness}

To realize a prescribed nonscalar direction $L$ with weight of order $t$, give it Kraus amplitude $\sqrt t$ and complete the remaining probability through $K_0$. With $B=L^\dagger L$, set
\begin{equation}
\label{app-robustness-eq-13}
K_1(t)=\sqrt t\,L,
\qquad
K_0(t)=\sqrt{I-tB},
\qquad
0\leq t\leq\|L\|_{\mathrm{op}}^{-2}.
\end{equation}

The bound on $t$ makes every eigenvalue of $I-tB$ nonnegative. The two terms $K_1^\dagger K_1=tB$ and $K_0^\dagger K_0=I-tB$ sum to $I$, so the channel $\mathcal E_t$ is exactly CPTP. To expand $K_0$, diagonalize the positive matrix $B$ and apply the Taylor expansion of $\sqrt{1-tb}$ to each eigenvalue $b$. This gives
\begin{equation}
\label{app-robustness-eq-14}
K_0(t)=I-\frac t2B-\frac{t^2}{8}B^2+O(t^3).
\end{equation}

For the loss, $K_1$ contributes $tQ_S(L)$. Projecting perpendicular to a probe removes the identity term of $K_0$, leaving $-t(I-P_i)B\psi_i/2+O(t^2)$. For the risk, the traceless part of $K_0$ is of order $t$ and its squared norm is of order $t^2$. At fixed $L$ and dimension this gives
\begin{equation}
\label{app-robustness-eq-15}
\begin{aligned}
\epsilon_S(\mathcal E_t)
&=tQ_S(L)+\frac{t^2}{4}Q_S(B)+O(t^3),\\
R(\mathcal E_t)
&=\frac{t}{d+1}
 \left\|L-\frac{\Tr\,L}{d}I\right\|_F^2+O(t^2).
\end{aligned}
\end{equation}

If $\mu_S>0$, choose a traceless minimizer $L$ of unit Frobenius norm. The leading terms are $\mu_St$ in the loss and $t/(d+1)$ in the risk, so
\begin{equation}
\label{app-robustness-eq-16}
\lim_{t\downarrow0}
\frac{R(\mathcal E_t)}{\epsilon_S(\mathcal E_t)}
=\frac{1}{(d+1)\mu_S}.
\end{equation}

The limit rules out a smaller coefficient. If $\mathcal A_S$ contains a nonscalar $L$, then $Q_S(L)=0$ and the loss is $O(t^2)$, while the risk has a positive linear term. This rules out any finite coefficient in the zero-gap case.

\subsubsection{Leading slope of the worst-case loss curve}
\label{app-robustness-what-the-optimal-coefficient-does-and-does-not-determine}

Define the worst-case curve
\begin{equation}
\label{app-robustness-eq-17}
W_S(\varepsilon)
=\sup\{R(\mathcal E):\mathcal E\ \text{CPTP},
\epsilon_S(\mathcal E)\leq\varepsilon\}.
\end{equation}

If $\mu_S>0$, the upper bound and the same minimizing family imply
\begin{equation}
\label{app-robustness-eq-18}
W_S(\varepsilon)=C_S\varepsilon+O(\varepsilon^2)
\qquad(\varepsilon\downarrow0),
\qquad
W_S(\varepsilon)\leq C_S\varepsilon\quad\text{for every }\varepsilon.
\end{equation}

The loss is analytic near zero with derivative $\mu_S>0$, so $\epsilon_S(\mathcal E_t)=\varepsilon$ has a small solution $t=\varepsilon/\mu_S+O(\varepsilon^2)$. Substituting into the risk gives a feasible channel of risk $C_S\varepsilon+O(\varepsilon^2)$. Equation~\eqref{app-robustness-eq-12} gives the upper bound.

\subsection{Exact identification}
\label{app-robustness-exact-identification}

Let the \textbf{overlap graph} have one vertex for each probe and an edge between distinct $i,j$ whenever $\langle\psi_i|\psi_j\rangle\neq0$.

\textbf{Known identification theorem.} The only CPTP channel with $\epsilon_S=0$ is the identity if and only if the probes span $\mathbb C^d$ and their overlap graph is connected. This is Theorem 1 of Mayer--Knill~\cite{mayer-knill-2018}. The minimum-setting consequences also appear in Li et al.~\cite{li-et-al-2021}.

At zero loss, Eq.~\eqref{app-robustness-eq-5} gives $K_a\psi_i=\lambda_{ai}\psi_i$ for every $a,i$. Taking diagonal and off-diagonal matrix elements of $\sum_aK_a^\dagger K_a=I$ between probes gives
\begin{equation}
\label{app-robustness-eq-19}
\sum_a|\lambda_{ai}|^2=1,
\qquad
\langle\psi_i|\psi_j\rangle
=\langle\psi_i|\psi_j\rangle
\sum_a\overline{\lambda_{ai}}\lambda_{aj}.
\end{equation}

The coefficient vectors $(\lambda_{ai})_a$ have unit norm and inner product one across each overlap edge, so they coincide there. Connectivity propagates this equality to every probe, and spanning makes each $K_a$ scalar on the whole space. The scalars have squared magnitudes summing to one, giving the identity channel.

If the probes occupy a proper subspace $V$, the unitary $I_V\oplus iI_{V^\perp}$ fixes every probe projector but changes the relative phase of a superposition across the two subspaces. If the overlap graph is disconnected, the spans of distinct components are orthogonal, so assigning distinct phases to two component spans has the same effect. In either case the nonscalar unitary has positive risk by Eq.~\eqref{app-robustness-eq-7}.

Exact identification constrains all Kraus operators jointly through trace preservation. The linear bound asks for a stronger condition on individual matrices in $\mathcal A_S$.

\subsection{Square-root upper bound for identifying ensembles}
\label{app-robustness-imported-square-root-upper-bound}

Mayer--Knill's square-root bound requires spanning probes and a connected overlap graph. Choose $d$ independent probes with Gram matrix $G_B$ and label them $1,\ldots,d$. Give each overlap edge $(i,j)$ length $|\langle\psi_i|\psi_j\rangle|^{-1/2}$. Let $W_{kl}$ be the shortest-path distance between basis vertices $k,l$, allowing paths through all probes. The coefficient is
\begin{equation}
\label{app-robustness-eq-20}
C_{\mathrm{MK}}
=\frac{2}{d}\left[
\frac{2}{d}\sum_{1\leq l< k\leq d}W_{kl}^2
+\sum_{k=1}^d\sqrt{(G_B^{-1})_{kk}}
\right].
\end{equation}

Mayer--Knill, Theorem 2 and Appendix B~\cite{mayer-knill-2018}, establish, with $s=\max_i e_i(\mathcal E)$,
\begin{equation}
\label{app-robustness-eq-21}
1-F_e(\mathcal E)
\leq C_{\mathrm{MK}}\sqrt s+O_S(s)
\qquad(s\downarrow0).
\end{equation}

The invertible Gram matrix and finite path lengths make $C_{\mathrm{MK}}$ finite. Absorbing the remainder into the square-root term gives $1-F_e\leq A_S\sqrt{s}$ for $s\leq s_0$, with $A_S,s_0>0$ fixed by the ensemble.

Set $p_{\min}=\min_i p_i>0$. The probe attaining $s$ contributes at least $p_{\min}s$ to the mean, so $s\leq\epsilon_S/p_{\min}$. Multiplying the bound on $1-F_e$ by $d/(d+1)$ gives the Haar risk bound
\begin{equation}
\label{app-robustness-eq-22}
R(\mathcal E)
\leq\frac{d A_S}{(d+1)\sqrt{p_{\min}}}
\sqrt{\epsilon_S(\mathcal E)}
\end{equation}

whenever $\epsilon_S\leq p_{\min}s_0$. Above that threshold, $R\leq1$ gives the bound with coefficient $(p_{\min}s_0)^{-1/2}$. Taking $B_S$ at least as large as both coefficients yields
\begin{equation}
\label{app-robustness-eq-23}
\boxed{
R(\mathcal E)\leq B_S\sqrt{\epsilon_S(\mathcal E)}
\quad\text{for every CPTP }\mathcal E.
}
\end{equation}

\subsection{Proof of Corollary~\ref{cor:classification}}
\label{app-robustness-the-sharp-robustness-trichotomy}

The nonidentifying case follows from the positive-risk channels constructed in Appendix~\ref{app-robustness-exact-identification}. For a positive gap, Appendix~\ref{app-robustness-what-the-optimal-coefficient-does-and-does-not-determine} gives the leading slope and quadratic remainder. For an identifying ensemble with zero gap, Appendix~\ref{app-robustness-imported-square-root-upper-bound} supplies the square-root upper bound.

To prove the matching square-root lower bound, choose a nonscalar $L\in\mathcal A_S$. If $Q_S(L^\dagger L)=0$, every probe would be an eigenvector of $L^\dagger L$. Distinct eigenvalues of a Hermitian matrix have orthogonal eigenspaces, so connectivity and spanning force $L^\dagger L=cI$. Here $c>0$ since $L\ne0$. Then $L/\sqrt c$ is a unitary that fixes all probes, and exact identification makes it scalar, a contradiction. Thus
\begin{equation}
\label{app-robustness-eq-24}
\beta_L=\frac14Q_S(L^\dagger L)>0,
\qquad
a_L=\frac{1}{d+1}
\left\|L-\frac{\Tr\,L}{d}I\right\|_F^2>0.
\end{equation}

For the exact family above,
\begin{equation}
\label{app-robustness-eq-25}
\epsilon_S(\mathcal E_t)=\beta_Lt^2+O(t^3),
\qquad
R(\mathcal E_t)=a_Lt+O(t^2).
\end{equation}

The analytic loss has positive derivative $2\beta_Lt+O(t^2)$ for small positive $t$, so it attains every sufficiently small $\varepsilon>0$. Taking a square root gives $\sqrt\varepsilon=\sqrt{\beta_L}\,t+O(t^2)$, or $t=\sqrt{\varepsilon/\beta_L}+O(\varepsilon)$. Substituting into the risk gives
\begin{equation}
\label{app-robustness-eq-26}
W_S(\varepsilon)
\geq\frac{a_L}{\sqrt{\beta_L}}\sqrt\varepsilon+O(\varepsilon).
\end{equation}

\subsection{First-order cancellation of survival loss}
\label{app-robustness-first-order-invisibility-is-a-statement-about-the-loss}

The jump Kraus term contributes $tLXL^\dagger$. Expanding $K_0XK_0^\dagger$ gives the two compensation terms $-tBX/2$ and $-tXB/2$. Together they yield
\begin{equation}
\label{app-robustness-eq-27}
\mathcal E_t(X)
=X+t\left(LXL^\dagger-\frac12\{L^\dagger L,X\}\right)+O(t^2).
\end{equation}

The derivative at $t=0$ has the standard single-jump form of a Gorini--Kossakowski--Sudarshan--Lindblad generator~\cite{gorini-et-al-1976,lindblad-1976}. When $L\psi_i=\lambda_i\psi_i$, the jump term on $P_i$ is $LP_iL^\dagger=|\lambda_i|^2P_i$. Also $P_iBP_i=\|L\psi_i\|^2P_i=|\lambda_i|^2P_i$. Splitting $BP_i$ and $P_iB$ at the projector shows how the portion along $P_i$ cancels, leaving
\begin{equation}
\label{app-robustness-eq-28}
\left.\frac{d}{dt}\mathcal E_t(P_i)\right|_{t=0}
=-\frac12\left[(I-P_i)BP_i+P_iB(I-P_i)\right].
\end{equation}

The expectation of this derivative in $P_i$ is zero. The off-diagonal terms can be nonzero, so the output can change at first order with survival loss beginning at second order. Kraus et al. show that pure stationary states require both the common jump-eigenvector condition and an additional effective-Hamiltonian condition (Theorem 1)~\cite{kraus-et-al-2008}.

Appendix~\ref{app-nonunitality-the-sharp-joint-error-frontier} computes the witness's nonunitality and shows how a bound on it limits the square-root regime.

\section{Probe geometry, exact witnesses, and sampling weights}
\label{app-probe-geometry}

Let $d\geq2$ and let $N$ distinct probe rays have positive weights. The variance gap $\mu_S$ controls the conversion from probe loss to Haar risk (Appendix~\ref{app-robustness}). The coordinate-support graph tests whether the gap is positive. We then compute two witnesses for zero-gap menus and optimize the weights for a positive-gap menu.

\subsection{A coordinate-support criterion}
\label{app-probe-geometry-a-coordinate-support-criterion}

Assume the probes span $\mathbb C^d$. Select a basis $b_1,\ldots,b_d$ from them, form the invertible matrix $B=[b_1\ \cdots\ b_d]$, and expand every remaining probe as
\begin{equation}
\label{app-probe-geometry-eq-1}
\psi_i=\sum_{j=1}^d c_{ij}b_j.
\end{equation}

The \textbf{coordinate-support graph} joins basis indices $j$ and $k$ whenever an extra probe has both $c_{ij}\neq0$ and $c_{ik}\neq0$. It has the same components on basis vertices as the vector matroid's dependence graph, in which each extra probe is joined to the basis vectors in its support~\cite{krogdahl-1977}.

\textbf{Coordinate-support proposition.} If this graph has $m$ connected components, then
\begin{equation}
\label{app-probe-geometry-eq-2}
\mathcal A_S
=\left\{B\mathrm{diag}(\lambda_1,\ldots,\lambda_d)B^{-1}:
\lambda_j\text{ is constant on each component}\right\},
\qquad
\dim_{\mathbb C}\mathcal A_S=m.
\end{equation}

In particular, $\mu_S>0$ if and only if the coordinate-support graph is connected.

\textbf{Proof.} Since $Ab_j=\lambda_j b_j$, the common right-eigenoperator $A$ has the stated diagonal representation even in a nonorthogonal basis. Matching the coefficient of $b_j$ in $A\psi_i=\lambda_i'\psi_i$ gives
\begin{equation}
\label{app-probe-geometry-eq-3}
(\lambda_j-\lambda_i')c_{ij}=0\qquad\text{for every }j.
\end{equation}

Every nonzero $c_{ij}$ forces $\lambda_j=\lambda_i'$, so eigenvalues agree along graph paths. Conversely, a single eigenvalue per component suffices because each extra probe has its entire support within one component. This gives $m$ free complex parameters. For $m=1$ the kernel is scalar. For $m>1$, subtracting $(\Tr A/d)I$ from a nonscalar kernel element and normalizing produces a traceless matrix of unit Frobenius norm with zero variance.

The component count is the dimension of $\mathcal A_S$ and is therefore independent of the chosen basis. Equivalently, $\mu_S>0$ exactly when there is no decomposition
\begin{equation}
\label{app-probe-geometry-eq-4}
\mathbb C^d=V_1\oplus\cdots\oplus V_m,\qquad m\geq2,
\end{equation}

with every ray contained in one summand. A disconnected graph supplies the summands by grouping basis vectors. Conversely, projection onto one summand along the others gives a nonscalar common right-eigenoperator. The summands may be nonorthogonal and the projections non-Hermitian.

\subsubsection{The two connectivity conditions}
\label{app-probe-geometry-the-two-connectivity-conditions}

The overlap-graph criterion is the Mayer--Knill identification theorem~\cite{mayer-knill-2018}, also used in Li et al.~\cite{li-et-al-2021}.

\begin{center}
\small
\setlength{\tabcolsep}{4pt}
\begin{tabular}{@{}p{0.26\linewidth}p{0.37\linewidth}p{0.3\linewidth}@{}}
\toprule
Graph & Vertices and edges & Property it detects \\
\midrule
Overlap graph & Two probe rays are joined if their inner product is nonzero & With spanning, exact identification of the unitary target \\
Coordinate-support graph & Two selected basis indices are joined when an extra probe uses both & A scalar common right-eigenoperator space and linear robustness \\
\bottomrule
\end{tabular}
\end{center}

A basis of $d$ probes has $d$ isolated vertices in the coordinate-support graph. A connected overlap graph still gives exact identification, though its zero variance gap puts arbitrary CPTP alternatives in the square-root regime. For example, $|0\rangle$ and $|+\rangle$ overlap and span, and a non-Hermitian matrix can assign them distinct eigenvalues. Adding $|1\rangle=\sqrt2|+\rangle-|0\rangle$ supplies the missing edge and forces those eigenvalues to agree. In any dimension, one extra ray with all basis coordinates nonzero joins every pair and restores a linear bound.

\subsection{An exact qubit example}
\label{app-probe-geometry-an-exact-qubit-example}

Take the equally weighted two-ray ensemble
\begin{equation}
\label{app-probe-geometry-eq-5}
S_2=\{|0\rangle,|+\rangle\},\qquad
|+\rangle=\frac{|0\rangle+|1\rangle}{\sqrt2}.
\end{equation}

Consider the jump matrix $L$ and its trace-preserving completion
\begin{equation}
\label{app-probe-geometry-eq-6}
L=\begin{pmatrix}0&1\\0&1\end{pmatrix},
\qquad
K_0=\begin{pmatrix}1&0\\0&r\end{pmatrix},
\qquad
K_1=\sqrt t\,L,
\qquad
r=\sqrt{1-2t},\quad0\leq t\leq\frac12.
\end{equation}

Here $K_1^\dagger K_1=2t|1\rangle\langle1|$, so $r^2=1-2t$ ensures trace preservation. Neither Kraus operator moves $|0\rangle$ out of its ray. On $|+\rangle$, the jump stays in the ray and only $K_0$ leaks, with amplitude $\langle -|K_0|+\rangle=(1-r)/2$ for $|-\rangle=(|0\rangle-|1\rangle)/\sqrt2$. On $|1\rangle$, only $K_1$ leaks, with amplitude $\langle 0|K_1|1\rangle=\sqrt{t}$. Squaring the amplitudes gives
\begin{equation}
\label{app-probe-geometry-eq-7}
e_0=0,
\qquad
e_+=\frac{(1-r)^2}{4},
\qquad
e_1=t.
\end{equation}

The fidelity identity in Appendix~\ref{app-robustness-two-exact-kraus-identities}, with Kraus traces $1+r$ and $\sqrt t$, gives $R=[4-(1+r)^2-t]/6$. Writing $h=1-r$ gives $2t=2h-h^2$ and $h=\sqrt{8\epsilon_{S_2}}$. Hence the exact curve on $0\leq t\leq1/2$ is
\begin{equation}
\label{app-probe-geometry-eq-8}
\epsilon_{S_2}=\frac{(1-r)^2}{8},
\qquad
R=\frac{1-r}{3}+\frac t6
=\sqrt{2\epsilon_{S_2}}-\frac23\epsilon_{S_2}.
\end{equation}

\subsubsection{Computing the gaps}
\label{app-probe-geometry-computing-the-gaps}

In the orthonormal Pauli basis write $A=(a_x\sigma_x+a_y\sigma_y+a_z\sigma_z)/\sqrt2$, with complex coefficients. Each ray has one perpendicular direction, so the leakage amplitudes are
\[
\begin{aligned}
\langle 1|A|0\rangle&=(a_x+\mathrm i a_y)/\sqrt2,\\
\langle -|A|+\rangle&=(-\mathrm i a_y+a_z)/\sqrt2,\\
\langle 0|A|1\rangle&=(a_x-\mathrm i a_y)/\sqrt2.
\end{aligned}
\]
Averaging the first two squared moduli gives $Q_{S_2}$. The two unit rows have overlap modulus $1/2$, so the nonzero spectrum comes from their Gram matrix scaled by $1/2$, with diagonal $1/2$ and off-diagonal modulus $1/4$.
\begin{equation}
\label{app-probe-geometry-eq-9}
\frac14
\begin{pmatrix}
1&i&0\\
-i&2&i\\
0&-i&1
\end{pmatrix},
\qquad
\mathrm{spec}(Q_{S_2})=\left\{0,\frac14,\frac34\right\}.
\end{equation}

Setting both amplitudes to zero gives $a_x=-\mathrm i a_y$ and $a_z=\mathrm i a_y$. This complex direction is spanned by the traceless part of $L$ in Eq.~\eqref{app-probe-geometry-eq-6}.

Give all three rays equal weights. The $a_x$ and $a_y$ cross terms for $|0\rangle$ and $|1\rangle$ cancel, leaving an $x$ eigenvalue of $1/3$ and a $2\times2$ block with trace $4/6$ and determinant $2/36$.
\begin{equation}
\label{app-probe-geometry-eq-10}
\frac16
\begin{pmatrix}
2&0&0\\
0&3&i\\
0&-i&1
\end{pmatrix},
\qquad
\mathrm{spec}(Q_{S_3})
=\left\{\frac{2-\sqrt2}{6},\frac13,
\frac{2+\sqrt2}{6}\right\}.
\end{equation}

Taking the smallest eigenvalue and applying the sharp coefficient $1/(3\mu_{S_3})$ gives
\begin{equation}
\label{app-probe-geometry-eq-11}
\mu_{S_3}=\frac{2-\sqrt2}{6},
\qquad
R\leq(2+\sqrt2)\epsilon_{S_3}.
\end{equation}

The commutator form (Appendix~\ref{app-nonunitality}) also includes leakage from the perpendicular ray into each probe. Half the sum of the two squared amplitudes is $(|a_x|^2+|a_y|^2)/2$ for $|0\rangle$ and $(|a_y|^2+|a_z|^2)/2$ for $|+\rangle$. Their average $(|a_x|^2+2|a_y|^2+|a_z|^2)/4$ has minimum $\gamma_{S_2}=1/4$ on traceless matrices of unit Frobenius norm, giving the sharp bound $R\leq4\epsilon_{S_2}/3$ for unital alternatives.

\subsection{A minimal identifying family in arbitrary dimension}
\label{app-probe-geometry-a-minimal-identifying-family-in-arbitrary-dimension}

Use the connected basis of Eq.~(15) in Ref.~\cite{li-et-al-2021},
\begin{equation}
\label{app-probe-geometry-eq-12}
u=\frac1{\sqrt d}\sum_{j=0}^{d-1}|j\rangle,
\qquad
S_-=\{|1\rangle,\ldots,|d-1\rangle,u\}.
\end{equation}

Only $u$ has a nonzero $|0\rangle$ coordinate, so it completes the other $d-1$ rays to a basis. Its nonzero overlaps with all those mutually orthogonal rays make the overlap graph a connected star. The basis therefore identifies the target exactly with $\mu_{S_-}=0$.

Choose
\begin{equation}
\label{app-probe-geometry-eq-13}
L=\sqrt d\,|u\rangle\langle0|,
\qquad
K_1=\sqrt t\,L,
\qquad
K_0=I+\left(\sqrt{1-dt}-1\right)|0\rangle\langle0|,
\qquad
0\leq t\leq\frac1d.
\end{equation}

Since $L^\dagger L=d|0\rangle\langle0|$, the channel is trace preserving. The computational probes in $S_-$ are fixed by $K_0$ and annihilated by $L$, and $Lu=u$. Put $h=1-\sqrt{1-dt}$ and let $b$ be the weight on $u$. Only $K_0u=u-(h/\sqrt d)|0\rangle$ contributes to the loss. The squared norm of the part of $|0\rangle$ perpendicular to $u$ is $1-1/d$, giving
\begin{equation}
\label{app-probe-geometry-eq-14}
e_u=\frac{d-1}{d^2}h^2,
\qquad
\epsilon_{S_-}=\frac{b(d-1)}{d^2}h^2.
\end{equation}

The Kraus traces $d-h$ and $\sqrt t$ give $R=(2dh-h^2-t)/[d(d+1)]$. Substituting $dt=2h-h^2$ and $h=d\sqrt{\epsilon_{S_-}}/\sqrt{b(d-1)}$ yields
\begin{equation}
\label{app-probe-geometry-eq-15}
\begin{aligned}
R
&=\frac{2(d-1)}{d^2}h
 -\frac{d-1}{d^2(d+1)}h^2\\
&=\frac{2\sqrt{d-1}}{d\sqrt b}\sqrt{\epsilon_{S_-}}
 -\frac{\epsilon_{S_-}}{b(d+1)}.
\end{aligned}
\end{equation}

For uniform weights, $b=1/d$, this becomes
\begin{equation}
\label{app-probe-geometry-eq-16}
R=2\sqrt{\frac{d-1}{d}}\sqrt{\epsilon_{S_-}}
-\frac{d}{d+1}\epsilon_{S_-}.
\end{equation}

Adding $|0\rangle$ detects the jump at first order. The output $K_1|0\rangle=\sqrt{dt}\,u$ gives loss $dt(1-1/d)=(d-1)t$. On the enlarged menu, $u$ has full support in the computational basis and makes the variance gap positive.

Both $R$ and $\epsilon_{S_-}$ increase with $h\in[0,1]$. At $h=1$ the mean loss with uniform weights is $(d-1)/d^3$ and the Haar risk is
\begin{equation}
\label{app-probe-geometry-eq-17}
R_{\max,\mathrm{witness}}
=\frac{(d-1)(2d+1)}{d^2(d+1)}=O(d^{-1}).
\end{equation}

The curve shows square-root behavior at fixed dimension, with a risk range that shrinks as $d$ grows. Appendix~\ref{app-product-channels} gives a construction at fixed global accuracy for independent qubit channels.

\subsection{The full basis plus one uniform ray}
\label{app-probe-geometry-the-full-basis-plus-one-uniform-ray}

Consider the basis-plus-superposition menu $S_+=\{|0\rangle,\ldots,|d-1\rangle,u\}$ studied in gate characterization~\cite{mayer-knill-2018}. Give each basis ray weight $a=(1-b)/d$ and the uniform ray weight $b\in(0,1)$.

Let $\Pi_{\mathrm{off}}$ be the Hilbert--Schmidt orthogonal projection onto matrices whose computational-basis diagonal vanishes, and let
\begin{equation}
\label{app-probe-geometry-eq-18}
\mathcal R_u(A)=(I-|u\rangle\langle u|)A|u\rangle\langle u|.
\end{equation}

The orthogonal projection $\mathcal R_u$ retains only the part of $A$ that maps $u$ into $u^\perp$, giving squared leakage $\|\mathcal R_u(A)\|_F^2$ on $u$. The squared leakage on $|j\rangle$ is $\sum_{i\ne j}|A_{ij}|^2$, so the sum over basis rays is $\|\Pi_{\mathrm{off}} A\|_F^2$. The variance operator is
\begin{equation}
\label{app-probe-geometry-eq-19}
Q_{S_+}=a\Pi_{\mathrm{off}}+b\mathcal R_u.
\end{equation}

\subsubsection{Spectrum}
\label{app-probe-geometry-spectrum}

An orthonormal basis $v_1,\ldots,v_{d-1}$ of $u^\perp$ gives an orthonormal basis $q_j=|v_j\rangle\langle u|$ for the range of $\mathcal R_u$. The traceless diagonal of $q_j$ has entries $(v_j)_k/\sqrt d$, so the squared norms of its diagonal and off-diagonal parts are $1/d$ and $(d-1)/d$. Normalize these two parts to give coordinates $(1/\sqrt d,\sqrt{(d-1)/d})$ for $q_j$. On the resulting plane, $\mathcal R_u$ projects onto $q_j$ and $\Pi_{\mathrm{off}}$ acts as $\mathrm{diag}(0,1)$. This gives
\begin{equation}
\label{app-probe-geometry-eq-20}
\begin{pmatrix}
b/d&b\sqrt{d-1}/d\\
b\sqrt{d-1}/d&a+b(d-1)/d
\end{pmatrix}.
\end{equation}

The diagonal parts inherit the orthogonality of the $v_j$. Since the full $q_j$ are also orthogonal, their off-diagonal parts are orthogonal as well. Each plane contributes two eigenvalues, determined by its trace $a+b$ and determinant $ab/d$.
\begin{equation}
\label{app-probe-geometry-eq-21}
\lambda_\pm
=\frac{a+b\pm\sqrt{(a+b)^2-4ab/d}}{2}.
\end{equation}

These planes use all $d-1$ traceless diagonal directions and $d-1$ of the $d(d-1)$ off-diagonal directions. The other $(d-1)^2$ off-diagonal directions are orthogonal to the $q_j$ and fixed by $\Pi_{\mathrm{off}}$, giving eigenvalue $a$. Both projections annihilate the identity. The spectrum is

\begin{center}
\small
\setlength{\tabcolsep}{4pt}
\begin{tabular}{@{}p{0.43\linewidth}p{0.52\linewidth}@{}}
\toprule
Eigenvalue & Multiplicity \\
\midrule
$0$ & $1$ \\
$a$ & $(d-1)^2$ \\
$\lambda_-$ & $d-1$ \\
$\lambda_+$ & $d-1$ \\
\bottomrule
\end{tabular}
\end{center}

The characteristic polynomial at $a$ is $-ab(1-1/d)<0$, placing $a$ between the two roots. Thus $\mu_{S_+}=\lambda_-$. Setting $a=b=1/(d+1)$ and rationalizing the reciprocal for $C_{S_+}$ gives
\begin{equation}
\label{app-probe-geometry-eq-22}
\mu_{S_+}
=\frac{1-\sqrt{1-1/d}}{d+1},
\qquad
C_{S_+}=d+\sqrt{d(d-1)}.
\end{equation}

\subsubsection{Optimizing all sampling weights}
\label{app-probe-geometry-optimizing-all-sampling-weights}

The gap is concave in the weights because it is a minimum of linear functions. Permuting the basis fixes $u$ and preserves the gap by unitary conjugation. Averaging the permuted weights therefore cannot lower the gap. The average assigns $(1-b)/d$ to each basis ray, so optimizing over $b$ gives the optimum over all distributions on the menu.

The smaller eigenvalue satisfies
\begin{equation}
\label{app-probe-geometry-eq-23}
\mu^2-\frac{1+(d-1)b}{d}\mu+\frac{b(1-b)}{d^2}=0.
\end{equation}

The gap vanishes at both endpoints and is positive and differentiable in the interior. At its interior maximum $\mathrm d\mu/\mathrm db=0$, so differentiating Eq.~\eqref{app-probe-geometry-eq-23} gives
\begin{equation}
\label{app-probe-geometry-eq-24}
\mu=\frac{1-2b}{d(d-1)}.
\end{equation}

Set $m=d-1$ and substitute $b=(1-dm\mu)/2$ into Eq.~\eqref{app-probe-geometry-eq-23} to get
\[
(m^2+4)(d\mu)^2-2(m+2)d\mu+1=0,
\]
with roots $d\mu=1/(m+2\pm2\sqrt m)$. The smaller block eigenvalue must lie below half the trace, so $2\mu-(a+b)=[(m^2+4)d\mu-(m+2)]/(2d)<0$. This selects the denominator $m+2+2\sqrt m$ and yields
\begin{equation}
\label{app-probe-geometry-eq-25}
\boxed{
b_*=\frac{1+\sqrt{d-1}}{d+1+2\sqrt{d-1}},
\qquad
\mu_*=\frac{1}{d[d+1+2\sqrt{d-1}]},
\qquad
C_*=\frac{d[d+1+2\sqrt{d-1}]}{d+1}.
}
\end{equation}

Concavity gives global optimality for this menu. At $d=2$, the optimal weights on $(|0\rangle,|+\rangle,|1\rangle)$ are $(3/10,4/10,3/10)$, giving $\mu_*=1/10$ and $C_*=10/3$. The optimized coefficient is $d+2\sqrt d+O(d^{-1/2})$ and uniform weighting gives an asymptotic coefficient of $2d$. Relative to the $d+1$-ray simplex with $C=d^2/(d+1)$, the optimized ratio is $1+(1+2\sqrt{d-1})/d$. Appendix~\ref{app-optimal-design} compares the ensembles attaining the universal support bound.

\subsection{Preparation and sampling resources}
\label{app-probe-geometry-preparation-and-sampling-resources}

When $d=2^n$, $S_-$ and $S_+$ contain $2^n$ and $2^n+1$ rays, respectively. To sample $S_+$ with weights $a,b$, choose $u$ with probability $b$ and otherwise draw $n$ uniform label bits. A basis input takes at most $n$ bit flips, and $u=|+\rangle^{\otimes n}$ takes $n$ Hadamards from $|0\rangle^{\otimes n}$. The computational basis and its image under a Hadamard on each qubit give a mutually unbiased pair with squared cross-overlaps $2^{-n}$. For the Mayer--Knill simplex~\cite{mayer-knill-2018}, Appendix~\ref{app-optimal-design-consequences-and-known-attaining-constructions} gives the product factors. Appendix~\ref{app-certification} gives the number of fresh outputs required for certification.

\section{Optimal support bound for arbitrary weighted ensembles}
\label{app-optimal-design}

This appendix proves Theorem~\ref{thm:design}. The proof starts from the rank of the projector moment. The slack in the rank bound forces balance at equality, and the moment spectrum then determines the weights and pairwise overlaps.

\subsection{Notation for the proof}
\label{app-optimal-design-notation-and-statement}

Let $S=\{(p_i,P_i)\}_{i=1}^N$ be a weighted ensemble of distinct normalized pure rays $P_i=|\psi_i\rangle\langle\psi_i|$ in dimension $d\geq2$, with $p_i>0$ and $\sum_i p_i=1$. Define
\begin{equation}
\label{app-optimal-design-eq-1}
\rho_S=\sum_i p_iP_i,
\qquad
Q_S(A)=\sum_i p_i\bigl\|(I-P_i)A|\psi_i\rangle\bigr\|^2,
\end{equation}

and the gap on the complex traceless operator space by
\begin{equation}
\label{app-optimal-design-eq-2}
\mu_S=
\min_{\substack{\Tr\,A=0\\\|A\|_F=1}}Q_S(A).
\end{equation}

When $\mu_S>0$, the global channel inequality holds with optimal coefficient
\begin{equation}
\label{app-optimal-design-eq-3}
R(\mathcal E)\leq C_S\epsilon_S(\mathcal E),
\qquad
C_S=\frac{1}{(d+1)\mu_S}.
\end{equation}

We use the weighted survival loss, Haar risk, and relative channel of Sec.~\ref{main-model}. Set $C_S=+\infty$ when no finite linear coefficient exists.

The equiangular-tight-frame condition used below is

\begin{equation}
\label{app-optimal-design-eq-6}
\frac1N\sum_iP_i=\frac{I}{d},
\qquad
\Tr(P_iP_j)=\frac{N-d}{d(N-1)}
\quad(i\neq j).
\end{equation}

Here $\Tr(P_iP_j)$ is the squared overlap between the rays. At $N=d^2$, the frame and projective-design characterizations agree.

\subsection{Operator representation}
\label{app-optimal-design-operator-representation}

Use column vectorization with the input factor first.
\begin{equation}
\label{app-optimal-design-eq-7}
|A\rangle\!\rangle
=\sum_{j,k}A_{jk}|k\rangle\otimes|j\rangle,
\qquad
\langle\!\langle A|B\rangle\!\rangle
=\Tr(A^\dagger B).
\end{equation}

Define operators on this $d^2$-dimensional Hilbert space by
\begin{equation}
\label{app-optimal-design-eq-8}
M=\sum_i p_i|P_i\rangle\!\rangle\langle\!\langle P_i|,
\qquad
\Pi_0=\frac{|I\rangle\!\rangle\langle\!\langle I|}{d},
\qquad
\mathsf Q=\rho_S^{\mathsf T}\otimes I-M.
\end{equation}

The column index is the first tensor factor, so $(\rho_S^{\mathsf T}\otimes I)|A\rangle\!\rangle=|A\rho_S\rangle\!\rangle$. The two terms in $\mathsf Q$ give
\[
\begin{aligned}
\langle\!\langle A|(\rho_S^{\mathsf T}\otimes I)|A\rangle\!\rangle
 &=\Tr(A^\dagger A\rho_S)=\sum_i p_i\|A|\psi_i\rangle\|^2,\\
\langle\!\langle A|M|A\rangle\!\rangle
 &=\sum_i p_i|\Tr(P_iA)|^2
 =\sum_i p_i|\langle\psi_i|A|\psi_i\rangle|^2 .
\end{aligned}
\]
The subtraction removes the component along the input ray, giving $Q_S(A)=\langle\!\langle A|\mathsf Q|A\rangle\!\rangle$. Since $Q_S$ is nonnegative and $Q_S(I)=0$, $\mathsf Q$ annihilates $|I\rangle\!\rangle$. The subspace perpendicular to this vector consists of traceless matrices, where the unit-sphere minimum gives
\begin{equation}
\label{app-optimal-design-eq-9}
\mathsf Q\succeq0,
\qquad
\mathsf Q|I\rangle\!\rangle=0,
\qquad
\mathsf Q\succeq\mu_S(I-\Pi_0).
\end{equation}

Each term in $M$ contributes trace $p_i$ because $\|P_i\|_F=1$. The range of $M$ lies in the span of the $N$ vectors $|P_i\rangle\!\rangle$. Since $\rho_S^{\mathsf T}\otimes I$ has trace $d$, the traces and rank on this operator space obey
\begin{equation}
\label{app-optimal-design-eq-10}
\Tr\,M=1,
\qquad
\mathrm{rank}\,M\leq N,
\qquad
\Tr\,\mathsf Q=d-1.
\end{equation}

To bound the gap by $\lambda_{\min}(\rho_S)$, choose a unit eigenvector $v$ for that eigenvalue and a unit vector $u\perp v$. The operator $A=|u\rangle\langle v|$ is traceless and has Frobenius norm one. It sends each probe to $|u\rangle\langle v|\psi_i\rangle$. Projecting off the input ray multiplies the squared norm of this vector by $1-|\langle\psi_i|u\rangle|^2$, so
\begin{equation}
\label{app-optimal-design-eq-11}
Q_S(A)
=\langle v|\rho_S|v\rangle
-\sum_i p_i|\langle\psi_i|u\rangle|^2
                  |\langle v|\psi_i\rangle|^2
\leq\lambda_{\min}(\rho_S).
\end{equation}

Thus $\mu_S\leq\lambda_{\min}(\rho_S)$. In particular, an ensemble that fails to span the ambient space has zero gap.

\subsection{Support bound for general mean states}
\label{app-optimal-design-bound-without-a-balance-assumption}

For a positive gap, choose $0<\mu<\mu_S$. The estimate $\mu_S\leq\lambda_{\min}(\rho_S)$ keeps the shifted mean $B$ positive definite.
\begin{equation}
\label{app-optimal-design-eq-12}
B=\rho_S-\mu I\succ0,
\qquad
H=B^{\mathsf T}\otimes I+\mu\Pi_0.
\end{equation}

Substituting $\mathsf Q=\rho_S^{\mathsf T}\otimes I-M$ into $\mathsf Q\succeq\mu(I-\Pi_0)$ gives $M\preceq H$. The operator $H$ retains the full mean state. Multiplying on both sides by $H^{-1/2}$ gives $0\preceq H^{-1/2}MH^{-1/2}\preceq I$. Invertibility preserves the rank, so at most $N$ nonzero eigenvalues contribute at most one each. By cyclicity, the trace of the rescaled moment is $\Tr(H^{-1}M)$, giving
\begin{equation}
\label{app-optimal-design-eq-13}
N\geq\Tr(H^{-1}M).
\end{equation}

Write
\begin{equation}
\label{app-optimal-design-eq-14}
s=\Tr\,B^{-1},
\qquad s_2=\Tr\,B^{-2},
\qquad
a=\Tr(\rho_SB^{-1})=d+\mu s,
\qquad
b=\Tr(\rho_SB^{-2})=s+\mu s_2.
\end{equation}

Write $D_B=B^{\mathsf T}\otimes I$ and $|w\rangle=|I\rangle\!\rangle/\sqrt d$. Then $H=D_B+\mu|w\rangle\langle w|$ and $\langle w|D_B^{-1}|w\rangle=s/d$. Inverting this rank-one update gives
\[
H^{-1}
=D_B^{-1}
-\frac{\mu D_B^{-1}|w\rangle\langle w|D_B^{-1}}{1+\mu s/d}.
\]
Here $D_B^{-1}|P_i\rangle\!\rangle=|P_iB^{-1}\rangle\!\rangle$. Since $P_i^2=P_i$, its diagonal element on $|P_i\rangle\!\rangle$ is $\Tr(B^{-1}P_i)$, averaging to $a$. The rank-one correction uses
\[
\langle w|D_B^{-1}|P_i\rangle\!\rangle
=\frac{\Tr(B^{-1}P_i)}{\sqrt d}
=\frac{\langle\psi_i|B^{-1}|\psi_i\rangle}{\sqrt d}.
\]
Taking the weighted sum of the diagonal elements of $H^{-1}$ on the vectors $|P_i\rangle\!\rangle$ and using $d(1+\mu s/d)=a$ gives
\begin{equation}
\label{app-optimal-design-eq-15}
\Tr(H^{-1}M)
=a-\frac{\mu}{a}
     \sum_i p_i\langle\psi_i|B^{-1}|\psi_i\rangle^2.
\end{equation}

Cauchy--Schwarz on $|\psi_i\rangle$ and $B^{-1}|\psi_i\rangle$, followed by the weighted mean, gives
\begin{equation}
\label{app-optimal-design-eq-16}
\sum_i p_i\langle\psi_i|B^{-1}|\psi_i\rangle^2
\leq\sum_i p_i\langle\psi_i|B^{-2}|\psi_i\rangle
=b,
\end{equation}
an upper bound on the sum subtracted in Eq.~\eqref{app-optimal-design-eq-15}.

Since $\mu/a>0$, the substitution bounds $\Tr(H^{-1}M)$ from below.
\begin{equation}
\label{app-optimal-design-eq-17}
N\geq F:=a-\frac{\mu b}{a}.
\end{equation}

The bound $F$ depends only on the eigenvalues $x_j=\lambda_j(\rho_S)-\mu>0$ of $B$. With $\tau=\sum_jx_j=1-d\mu$, $s=\sum_j x_j^{-1}$, and $s_2=\sum_j x_j^{-2}$, we use
\begin{equation}
\label{app-optimal-design-eq-18}
\tau s\geq d^2,
\qquad
\tau(s^2-s_2)\geq d(d-1)s.
\end{equation}

The first follows by applying Cauchy--Schwarz to the vectors with entries $\sqrt{x_j}$ and $1/\sqrt{x_j}$. Their inner product is $d$, and equality holds exactly when all $x_j$ coincide.

For the second inequality, start from $s^2-s_2=2\sum_{i<j}(x_i x_j)^{-1}$ and multiply by $\tau=\sum_k x_k$. The terms in which $k$ equals $i$ or $j$ contribute $2(d-1)s$. For each triple of distinct indices the three choices of the numerator give the second term on the first line below.
\begin{equation}
\label{app-optimal-design-eq-19}
\begin{aligned}
\tau(s^2-s_2)
&=2(d-1)s
 +2\sum_{i< j< k}
   \frac{x_i^2+x_j^2+x_k^2}{x_ix_jx_k}\\
&\geq2(d-1)s
 +2\sum_{i< j< k}
   \left(\frac1{x_i}+\frac1{x_j}+\frac1{x_k}\right)\\
&=d(d-1)s.
\end{aligned}
\end{equation}

The inequality for each triple follows from the sum of its three squared pairwise differences. Each reciprocal occurs in $\binom{d-1}{2}$ triples, giving the final factor $d(d-1)$. For $d=2$ the triple terms are absent and the inequality is an identity.

To use these estimates, write the difference between $F$ and the proposed trace bound over the common denominator $a\tau$. The numerator is $\tau(a^2-\mu b)-ad(1-\mu)$. Substitute $a=d+\mu s$ and $b=s+\mu s_2$, then apply the second inequality of Eq.~\eqref{app-optimal-design-eq-18} to the term containing $s^2-s_2$.
\begin{equation}
\label{app-optimal-design-eq-20}
\begin{aligned}
a\tau\left(F-\frac{d(1-\mu)}{1-d\mu}\right)
&=(d-1)\mu\bigl[-d^2+(1-2d\mu)s\bigr]
  +\tau\mu^2(s^2-s_2)\\
&\geq(d-1)\mu(\tau s-d^2)\\
&\geq0.
\end{aligned}
\end{equation}

Therefore
\begin{equation}
\label{app-optimal-design-eq-21}
N\geq\frac{d(1-\mu)}{1-d\mu}.
\end{equation}

The right-hand side exceeds $d$ by $d(d-1)\mu/(1-d\mu)>0$. This rules out a positive gap for $N\leq d$. For $N>d$, multiplying by $1-d\mu>0$ and collecting the terms proportional to $\mu$ gives $d(N-1)\mu\leq N-d$. Taking $\mu\uparrow\mu_S$ gives
\begin{equation}
\label{app-optimal-design-eq-22}
\mu_S\leq\frac{N-d}{d(N-1)}.
\end{equation}

\subsection{Equality and the potentially singular endpoint}
\label{app-optimal-design-equality-and-the-potentially-singular-endpoint}

Suppose $d< N\leq d^2$ and the cardinality bound is saturated. Define
\begin{equation}
\label{app-optimal-design-eq-23}
q=\frac{N-d}{d(N-1)}=\mu_S>0.
\end{equation}

At $\mu=q$, equality in $\lambda_{\min}(\rho_S)\geq q$ would make $B$ singular. Since $1-dq=(d-1)/(N-1)>0$, the leading term in the trace bound has a finite limit. To track the remaining contribution, retain the slack from Eq.~\eqref{app-optimal-design-eq-20}.
\begin{equation}
\label{app-optimal-design-eq-24}
F-\frac{d(1-\mu)}{1-d\mu}
\geq
\frac{(d-1)\mu(\tau s-d^2)}{(d+\mu s)\tau}.
\end{equation}

If $\lambda_{\min}(\rho_S)=q$, at least one term of $s=\sum_j[\lambda_j(\rho_S)-\mu]^{-1}$ diverges as $\mu\uparrow q$. Factor the slack as
\[
(d-1)\frac{\mu s}{d+\mu s}
\left(1-\frac{d^2}{\tau s}\right).
\]
Both factors after $d-1$ tend to one because $q>0$ and $\tau\to1-dq>0$. Since the leading term tends to $d(1-q)/(1-dq)=N$, the lower bound on $F$ tends to $N+d-1$, contradicting $N\geq F$. Equality requires $\lambda_{\min}(\rho_S)>q$, so $B$ remains invertible at the endpoint.

The bound is then continuous up to $\mu=q$. Since $N=d(1-q)/(1-dq)$, the slack must vanish, so $\tau s=d^2$ there. The Cauchy--Schwarz equality condition in Eq.~\eqref{app-optimal-design-eq-18} makes the $x_j$, and hence the eigenvalues of $\rho_S$, equal. Taking the trace yields
\begin{equation}
\label{app-optimal-design-eq-25}
\rho_S=I/d.
\end{equation}

Since each $P_i$ has trace one, $M|I\rangle\!\rangle=|\rho_S\rangle\!\rangle=|I\rangle\!\rangle/d$. Hermiticity then makes the perpendicular traceless subspace invariant under $M$. On that subspace $\mathsf Q=I/d-M$ has eigenvalues at least $q$, so the eigenvalues of $M$ are at most
\begin{equation}
\label{app-optimal-design-eq-26}
c=\frac1d-q=\frac{d-1}{d(N-1)}.
\end{equation}

Removing the identity direction leaves trace $(d-1)/d=(N-1)c$ and rank at most $N-1$. Since each eigenvalue is bounded by $c$, the traceless part has exactly $N-1$ nonzero eigenvalues, all equal to $c$.

Consider the weighted projector Gram matrix
\begin{equation}
\label{app-optimal-design-eq-27}
G_{ij}=\sqrt{p_ip_j}\Tr(P_iP_j),
\qquad
v=(\sqrt{p_1},\ldots,\sqrt{p_N})^{\mathsf T}.
\end{equation}

Let $V$ have columns $\sqrt{p_i}|P_i\rangle\!\rangle$, so $M=VV^\dagger$ and $G=V^\dagger V$. Since $MV=VG$, $V$ maps the eigenspace of $G$ at $\lambda>0$ to that of $M$, with inverse $V^\dagger/\lambda$. Thus the nonzero spectra agree with multiplicities. Also $V^\dagger|I\rangle\!\rangle=v$ and $Vv=|\rho_S\rangle\!\rangle$. Balance gives $Gv=v/d$, and $v$ is a unit vector because $\sum_i p_i=1$. The spectral decomposition is
\begin{equation}
\label{app-optimal-design-eq-28}
G=cI_N+\left(\frac1d-c\right)vv^\dagger.
\end{equation}

Since $P_i^2=P_i$, the diagonal entry $G_{ii}$ equals $p_i$. Comparing diagonal entries gives
\begin{equation}
\label{app-optimal-design-eq-29}
p_i=c+\left(\frac1d-c\right)p_i,
\qquad\text{hence}\qquad p_i=\frac1N.
\end{equation}

For distinct indices, the off-diagonal entries in Eq.~\eqref{app-optimal-design-eq-28} are $(1/d-c)\sqrt{p_i p_j}$. Dividing by the positive factor $\sqrt{p_i p_j}$ in Eq.~\eqref{app-optimal-design-eq-27} gives
\begin{equation}
\label{app-optimal-design-eq-30}
\Tr(P_iP_j)=\frac1d-c
=\frac{N-d}{d(N-1)}.
\end{equation}

Conversely, a uniformly weighted equiangular tight frame is balanced. Its Gram matrix has diagonal entries $1/N$ and off-diagonal entries $q/N$. The normalized all-ones vector has eigenvalue $[1+(N-1)q]/N=1/d$, and its orthogonal complement has eigenvalue $(1-q)/N=c$. Therefore the smallest traceless eigenvalue of $I/d-M$ is $1/d-c=q$.

\subsection{Absolute optimum and projective designs}
\label{app-optimal-design-absolute-optimum-and-projective-designs}

The trace $d-1$ of $\mathsf Q$ lies entirely on the traceless subspace. Its smallest eigenvalue there is at most the average.
\begin{equation}
\label{app-optimal-design-eq-31}
\mu_S\leq\frac{d-1}{d^2-1}=\frac1{d+1}.
\end{equation}

Equality requires every traceless eigenvalue to equal this average, so
\begin{equation}
\label{app-optimal-design-eq-32}
\mathsf Q=\frac{I-\Pi_0}{d+1}.
\end{equation}

To derive balance from this operator equation, write $|P_i\rangle\!\rangle=|\overline{\psi_i}\rangle\otimes|\psi_i\rangle$, where the bar denotes complex conjugation in the chosen basis. Each projector term in $M$ is then $P_i^{\mathsf T}\otimes P_i$. Taking the trace over the output factor therefore sends $M$ to $\rho_S^{\mathsf T}$ and $\rho_S^{\mathsf T}\otimes I$ to $d\rho_S^{\mathsf T}$. The partial trace of $\Pi_0$ is $I/d$, as follows by tracing the second factor of $\sum_{j,k}|j\rangle\langle k|\otimes|j\rangle\langle k|/d$. Applying these rules to Eq.~\eqref{app-optimal-design-eq-32} gives
\[
(d-1)\rho_S^{\mathsf T}=\frac{dI-I/d}{d+1}=\frac{d-1}{d}I.
\]
Hence $\rho_S=I/d$. Substituting this into $\mathsf Q=I/d-M$ and using Eq.~\eqref{app-optimal-design-eq-32} yields
\begin{equation}
\label{app-optimal-design-eq-33}
M=\frac{I+|I\rangle\!\rangle\langle\!\langle I|}{d(d+1)}.
\end{equation}

Transposing the input factor of $P_i^{\mathsf T}\otimes P_i$ gives $P_i\otimes P_i$. On the right-hand side of Eq.~\eqref{app-optimal-design-eq-33}, the identity is unchanged, and the input transpose sends
\[
|I\rangle\!\rangle\langle\!\langle I|
=\sum_{j,k}|j\rangle\langle k|\otimes|j\rangle\langle k|
\quad\longmapsto\quad
\sum_{j,k}|k\rangle\langle j|\otimes|j\rangle\langle k|=\mathbb F.
\]
The resulting state moment is
\begin{equation}
\label{app-optimal-design-eq-34}
\sum_i p_iP_i\otimes P_i
=\frac{I+\mathbb F}{d(d+1)},
\end{equation}

where $\mathbb F$ exchanges the two factors. The right-hand side is the Haar second moment, so this is the weighted complex projective $2$-design condition. Conversely, taking the output trace gives $\rho_S=I/d$ since $\Tr_{\rm out}\mathbb F=I$. Transposing the input factor gives Eq.~\eqref{app-optimal-design-eq-33}, and subtracting it from $I/d$ recovers Eq.~\eqref{app-optimal-design-eq-32}. The positive identity term in Eq.~\eqref{app-optimal-design-eq-33} gives $M$ full rank $d^2$, so the rank bound requires $N\geq d^2$. At $N=d^2$ the attaining ETF gives a symmetric informationally complete measurement~\cite{renes-et-al-2004}. Weighted unions of orthonormal bases provide other constructions of projective $2$-designs~\cite{roy-scott-2007}.

\subsection{Optimizing weights on a prescribed menu}
\label{app-optimal-design-optimizing-weights-on-a-prescribed-menu}

Use the sampling simplex $\Delta_N$ and the single-probe forms of Appendix~\ref{app-certification-geometry-and-risk-identities}. Let $\mathsf Q_i$ represent each form on traceless matrices. The best gap and coefficient are
\begin{equation}
\label{app-optimal-design-eq-35}
\mu_*=
\max_{p\in\Delta_N}
\lambda_{\min}\left(\left.\sum_i p_i\mathsf Q_i\right|_{\Tr\,A=0}\right),
\qquad
C_*=\frac1{(d+1)\mu_*}
\quad(\mu_*>0).
\end{equation}

The gap is concave in $p$ because it is a minimum of linear functions. Each probe form takes values in $[0,1]$ on the traceless unit sphere, so the gap changes by at most $\sum_i|p_i-p_i'|$. It therefore attains its maximum on the compact simplex. Appendix~\ref{app-certification-geometry-and-risk-identities} explains why allowing zero weights leaves the supremum over positive weights unchanged.

The support theorem counts the $m$ rays of positive weight. The gap vanishes for $m\leq d$. For $d<m\leq d^2$, increasing the support size from $m$ to $m+1$ within this range raises the upper bound by $(d-1)/[dm(m-1)]>0$. Hence attaining the bound for an $N$-ray menu with $d<N\leq d^2$ requires all $N$ rays in the support and uniform weights on an equiangular tight frame. For $N\geq d^2$, $C_*=1$ exactly when the menu admits a weighted projective $2$-design, possibly on a smaller support.

Appendix~\ref{app-certification-exact-optimized-sampling-coefficient} converts this optimized coefficient into the leading cost of adaptive certification by using the common channel constructed in Appendix~\ref{app-certification-exact-multi-kraus-adversary}.

\subsection{Consequences and known attaining constructions}
\label{app-optimal-design-consequences-and-known-attaining-constructions}

Writing $N=d+k$ gives
\begin{equation}
\label{app-optimal-design-eq-36}
C_*\geq\frac{d(d+k-1)}{(d+1)k}
\qquad(1\leq k\leq d^2-d).
\end{equation}

For $k\ll d$, the bound is of order $d/k$. One extra ray can change the precision exponent, with coefficient still at least $d^2/(d+1)$. A dimension-independent coefficient requires $\Omega(d)$ extra rays. Two mutually unbiased bases suffice with $d$ extra rays and a coefficient below $2$.

At support size $d+1$, the uniform simplex attains the bound. Its rays can be chosen as
\begin{equation}
 \ket{\phi_k}=\frac1{\sqrt d}\sum_{j=0}^{d-1}
 e^{2\pi\mathrm i jk/(d+1)}\ket j,\qquad k=0,\ldots,d.
 \label{app-optimal-design-eq-simplex}
\end{equation}
Let $\omega=e^{2\pi\mathrm i/(d+1)}$. The $(j,j')$ entry of the mean projector is $(1/[d(d+1)])\sum_{k=0}^d\omega^{(j-j')k}$. This sum is $d+1$ for $j=j'$ and zero otherwise, giving $I/d$. For $k\neq k'$, the full series $\sum_{j=0}^d\omega^{j(k'-k)}$ vanishes. The overlap in Eq.~\eqref{app-optimal-design-eq-simplex} omits the last term, so its magnitude is $1/d$. The squared overlap $1/d^2$ is the value of Eq.~\eqref{app-optimal-design-eq-6} at $N=d+1$.

For $d=2^n$, write $j=\sum_{a=0}^{n-1}2^a j_a$ and order the qubits by the bit index $a$. The amplitude phase factors as $\omega^{jk}=\prod_a\omega^{2^a j_a k}$, so each ray is the product
\[
\ket{\phi_k}
=\bigotimes_{a=0}^{n-1}\frac{\ket0+\omega^{2^ak}\ket1}{\sqrt2}.
\]
The menu has $d+1$ entries, exponential in $n$, and one ray is sampled per call.

For two equally sampled mutually unbiased bases, the contribution of each basis to $M$ is the projector onto its diagonal operator subspace, scaled by $1/(2d)$. Indeed, the rank-one projectors of an orthonormal basis are orthonormal in the Frobenius inner product. The two diagonal subspaces both contain the identity. Their traceless parts are orthogonal, since for one projector $P$ from the first basis and another $P'$ from the second, $\Tr[(P-I/d)(P'-I/d)]=\Tr(PP')-1/d=0$. Thus $M$ has eigenvalue $1/(2d)$ on their traceless parts and is zero on the remaining traceless directions. The ensemble is balanced, so subtracting the largest traceless eigenvalue of $M$ from $1/d$ gives $\mu_S=1/(2d)$.

\begin{center}
\small
\setlength{\tabcolsep}{4pt}
\begin{tabular}{@{}p{0.35\linewidth}p{0.14\linewidth}p{0.2\linewidth}p{0.24\linewidth}@{}}
\toprule
Ensemble & Support size & Gap $\mu_S$ & Coefficient $C_S$ \\
\midrule
Uniform equiangular tight frame, when it exists & $N$ & $(N-d)/[d(N-1)]$ & $d(N-1)/[(d+1)(N-d)]$ \\
Uniform simplex frame & $d+1$ & $1/d^2$ & $d^2/(d+1)$ \\
Two uniformly sampled mutually unbiased bases & $2d$ & $1/(2d)$ & $2d/(d+1)$ \\
Weighted complex projective $2$-design & At least $d^2$ & $1/(d+1)$ & $1$ \\
\bottomrule
\end{tabular}
\end{center}

Mayer and Knill gave the symmetric-frame constructions and their fidelity coefficients~\cite{mayer-knill-2018}. Zhu and Zhang formulated process verification for arbitrary weighted ensembles and derived explicit spectral bounds for balanced ensembles~\cite{zhu-zhang-2020}. The rank argument above gives the support-size bound on the global linear coefficient and derives balance at equality.

A verifier can sample from a large menu without visiting every ray. Two $d$-element bases need an order-$\log d$ classical label. The coefficient in Appendix~\ref{app-certification-exact-optimized-sampling-coefficient} keeps $d$ and the menu fixed and counts fresh channel uses. Preparation and target measurements have their own costs.

\section{Unitality and the nonunitality frontier}
\label{app-nonunitality}

This appendix proves Proposition~\ref{prop:noise}. The commutator form isolates the correction to survival loss when a channel moves the maximally mixed state. We use it to derive the sharp unital bound and determine how the risk changes when that displacement is bounded.

The relative channel $\mathcal E=\mathcal U^\dagger\circ\mathcal V$ is unital exactly when $\mathcal V$ is unital, since the target $\mathcal U$ is unitary and fixes $I$. We assume that the pure probes span $\mathbb C^d$, their overlap graph is connected, and all weights are positive.

\subsection{The commutator form}
\label{app-nonunitality-the-commutator-form}

Let $\overline\rho=\rho_S=\sum_i p_iP_i$ be the ensemble's mean input state and let $[A,P]=AP-PA$. Define
\begin{equation}
\label{app-nonunitality-eq-1}
G_S(A)=\frac12\sum_i p_i\|[A,P_i]\|_F^2,
\qquad
\gamma_S=
\min_{\substack{\Tr\,A=0\\\|A\|_F=1}}G_S(A).
\end{equation}

The minimization runs over the full complex traceless matrix space. Decomposing the space into the probe ray and its orthogonal complement exposes the two blocks of $A$ measured by the commutator,
\[
[A,P_i]=(I-P_i)AP_i-P_iA(I-P_i).
\]
The blocks map between the two subspaces in opposite directions and are Hilbert--Schmidt orthogonal. Taking the adjoint of $P_iA(I-P_i)$ shows that their squared norms are $\|(I-P_i)A\psi_i\|^2$ and $\|(I-P_i)A^\dagger\psi_i\|^2$. Expanding gives
\begin{equation}
\label{app-nonunitality-eq-2}
\|[A,P_i]\|_F^2
=\langle\psi_i|A^\dagger A|\psi_i\rangle
+\langle\psi_i|AA^\dagger|\psi_i\rangle
-2|\langle\psi_i|A|\psi_i\rangle|^2.
\end{equation}

Averaging the two probe variances over the ensemble gives
\begin{equation}
\label{app-nonunitality-eq-3}
G_S(A)=\frac12[Q_S(A)+Q_S(A^\dagger)].
\end{equation}

With positive weights, $G_S(A)=0$ requires $AP_i=P_iA$ for every probe. Taking the column and row along $\psi_i$ gives $A\psi_i=\lambda_i\psi_i$ and $\langle\psi_i|A=\lambda_i\langle\psi_i|$, with $\lambda_i=\langle\psi_i|A|\psi_i\rangle$. Across a nonzero overlap,
\begin{equation}
\label{app-nonunitality-eq-4}
\lambda_i\langle\psi_i|\psi_j\rangle
=\langle\psi_i|A|\psi_j\rangle
=\lambda_j\langle\psi_i|\psi_j\rangle,
\end{equation}

so the two eigenvalues coincide. Connectivity propagates this equality along paths, and spanning gives $A=\lambda I$ on the full space. The traceless Frobenius-unit sphere is compact and contains no scalar matrix, so the minimum $\gamma_S$ is positive.

If the probes do not span, project onto their span. If the overlap graph is disconnected, project onto one component span, which is orthogonal to the other component spans. Either projector is nonscalar and commutes with all $P_i$. Its normalized traceless part gives $\gamma_S=0$.

\subsection{The exact channel identity}
\label{app-nonunitality-the-exact-channel-identity}

Commutator squares also enter the characterization of fixed points of trace-preserving unital maps by the commutant of their Kraus operators~\cite{arias-et-al-2002}.

Trace preservation and unitality enter separately when Eq.~\eqref{app-nonunitality-eq-2} is summed over the Kraus operators. Since $\sum_a K_a^\dagger K_a=I$, the first term sums to one. The second sums to $\langle\psi_i|\mathcal E(I)|\psi_i\rangle$, using $\mathcal E(I)=\sum_a K_aK_a^\dagger$. The squared matrix elements sum to the survival probability $1-e_i(\mathcal E)$. Hence
\[
\frac12\sum_a\|[K_a,P_i]\|_F^2
=e_i(\mathcal E)+\frac12\langle\psi_i|\mathcal E(I)-I|\psi_i\rangle .
\]
Taking the weighted mean gives the channel identity
\begin{equation}
\label{app-nonunitality-eq-5}
\boxed{
\sum_aG_S(K_a)
=\epsilon_S(\mathcal E)
+\frac12\Tr
\left[\overline\rho\bigl(\mathcal E(I)-I\bigr)\right].
}
\end{equation}

The scalar part of a Kraus operator commutes with each projector. Set $\widetilde K_a=K_a-(\Tr K_a/d)I$. Homogeneity and the gap bound give $G_S(K_a)=G_S(\widetilde K_a)\geq\gamma_S\|\widetilde K_a\|_F^2$. Summing and using the Haar-risk identity in Eq.~\eqref{app-robustness-eq-7} gives
\[
\sum_a G_S(K_a)\geq\gamma_S\sum_a\|\widetilde K_a\|_F^2
=(d+1)\gamma_S R(\mathcal E).
\]
For a unital channel, Eq.~\eqref{app-nonunitality-eq-5} makes the left-hand side exactly the measured loss, giving
\begin{equation}
\label{app-nonunitality-eq-6}
\boxed{
R(\mathcal E)\leq
\frac{\epsilon_S(\mathcal E)}{(d+1)\gamma_S}
\qquad\text{if }\mathcal E(I)=I.
}
\end{equation}

The bound applies even when $\mu_S=0$. General process-verification bounds and their ensemble assumptions are compared in Zhu--Zhang~\cite{zhu-zhang-2020}.

\subsubsection{Sharpness within the unital class}
\label{app-nonunitality-sharpness-within-the-unital-class}

To find a direction that attains the coefficient in the small-error limit, write a traceless matrix as $A=X+iY$ with $X=(A+A^\dagger)/2$ and $Y=(A-A^\dagger)/(2i)$. Both parts are traceless and Hermitian. The Hilbert--Schmidt inner product is real for two Hermitian matrices or two anti-Hermitian matrices, such as $[X,P_i]$ and $[Y,P_i]$. The cross terms involving $i$ therefore vanish in both quadratic forms, giving
\begin{equation}
\label{app-nonunitality-eq-7}
\|A\|_F^2=\|X\|_F^2+\|Y\|_F^2,
\qquad
G_S(A)=G_S(X)+G_S(Y).
\end{equation}

The quotient $G_S(A)/\|A\|_F^2$ is a weighted mean of the quotients of the nonzero Hermitian parts, with weights proportional to their squared norms. A nonzero part of a minimizer must therefore attain the same minimum. Normalizing it gives a traceless Hermitian minimizer $H$ with $\|H\|_F=1$.

Use the unitary channel $\mathcal V_s(X)=V_sXV_s^\dagger$ with $V_s=e^{isH}$. The variance for the loss uses $V_s-I$, and the Haar-risk identity uses $V_s-(\Tr V_s/d)I$. Since $H$ is traceless, both matrix arguments equal $isH+O(s^2)$. The forms are quadratic in these arguments, and $Q_S(H)=G_S(H)$, giving
\begin{equation}
\label{app-nonunitality-eq-8}
\epsilon_S(\mathcal V_s)=s^2Q_S(H)+O(s^3)
=s^2\gamma_S+O(s^3),
\qquad
R(\mathcal V_s)=\frac{s^2}{d+1}+O(s^3).
\end{equation}

The ratio of risk to survival infidelity as $s\to0$ proves sharpness.

\subsection{Quantitative deviation from unitality}
\label{app-nonunitality-quantitative-deviation-from-unitality}

Measure the change in the maximally mixed state by trace distance, with $\|X\|_1=\Tr\sqrt{X^\dagger X}$ denoting the trace norm.
\begin{equation}
\label{app-nonunitality-eq-9}
u(\mathcal E)
=\frac12\left\|\mathcal E(I/d)-I/d\right\|_1.
\end{equation}

Set $\Delta=\mathcal E(I)-I$, so $\|\Delta\|_1=2d\,u(\mathcal E)$. Trace preservation gives $\Tr\,\Delta=0$, so centering $\overline\rho$ at the midpoint $m=[\lambda_{\max}(\overline\rho)+\lambda_{\min}(\overline\rho)]/2$ leaves $\Tr(\overline\rho\Delta)$ unchanged. The norm $\|\overline\rho-mI\|_{\mathrm{op}}$ is then half the spectral range. The trace inequality
\[
|\Tr(\overline\rho\Delta)|
=|\Tr[(\overline\rho-m I)\Delta]|
\leq\|\overline\rho-m I\|_{\mathrm{op}}\|\Delta\|_1
\]
then gives
\begin{equation}
\label{app-nonunitality-eq-10}
\begin{aligned}
\Tr(\overline\rho\Delta)
&\leq\frac{\lambda_{\max}(\overline\rho)
-\lambda_{\min}(\overline\rho)}{2}\|\Delta\|_1\\
&=d\left[\lambda_{\max}(\overline\rho)
-\lambda_{\min}(\overline\rho)\right]u(\mathcal E).
\end{aligned}
\end{equation}

Combining this with the exact channel identity and the commutator gap proves
\begin{equation}
\label{app-nonunitality-eq-11}
\boxed{
R(\mathcal E)\leq
\frac{\epsilon_S(\mathcal E)
+\dfrac d2\left[\lambda_{\max}(\overline\rho)
-\lambda_{\min}(\overline\rho)\right]u(\mathcal E)}
{(d+1)\gamma_S}.
}
\end{equation}

The inequality holds for every CPTP alternative. For a balanced ensemble, $\overline\rho=I/d$, the correction vanishes for every channel. The difference $Q_S(A)-Q_S(A^\dagger)=\Tr[\overline\rho(A^\dagger A-AA^\dagger)]$ also vanishes by cyclicity of the trace. Equation~\eqref{app-nonunitality-eq-3} then gives $Q_S=G_S$, so $\mu_S=\gamma_S$. An identifying ensemble with $\mu_S=0$ must therefore be unbalanced.

\subsection{The sharp joint error frontier}
\label{app-nonunitality-the-sharp-joint-error-frontier}

For a permitted training loss $\varepsilon$ and a physical nonunitality budget $\eta$, define
\begin{equation}
\label{app-nonunitality-eq-12}
\Phi_S(\varepsilon,\eta)
=\sup\left\{R(\mathcal E):
\mathcal E\ \text{CPTP},\quad
\epsilon_S(\mathcal E)\leq\varepsilon,\quad
u(\mathcal E)\leq\eta\right\}.
\end{equation}

For a connected, spanning ensemble with $\mu_S=0$, the frontier in Proposition~\ref{prop:noise} is
\begin{equation}
\label{app-nonunitality-eq-13}
\boxed{
\Phi_S(\varepsilon,\eta)
=\Theta_S\!\left(\varepsilon+
\min\{\eta,\sqrt\varepsilon\}\right).
}
\end{equation}

The lower and upper comparison constants $c_S,C_S'>0$ are valid for $0\leq\varepsilon\leq\varepsilon_0$ and all $\eta\geq0$. These constants and $\varepsilon_0>0$ depend only on the ensemble and dimension.

\textbf{Upper bound.} Substituting both budgets in Eq.~\eqref{app-nonunitality-eq-11} and taking the larger coefficient gives $R\leq A_S(\varepsilon+\eta)$ for a finite $A_S$. The Mayer--Knill bound gives $R\leq B_S\sqrt\varepsilon$ independently of $\eta$ (Appendix~\ref{app-robustness-imported-square-root-upper-bound}). Taking the better bound gives $\max\{A_S,B_S\}$ times
\begin{equation}
\label{app-nonunitality-eq-14}
\min\{\varepsilon+\eta,\sqrt\varepsilon\}
\leq\varepsilon+\min\{\eta,\sqrt\varepsilon\}.
\end{equation}

\textbf{Lower bound.} A depolarizing channel uses the loss allowance while leaving the maximally mixed state unchanged. Take
\begin{equation}
\label{app-nonunitality-eq-15}
\mathcal D_q(X)=(1-q)X+q\Tr(X)I/d
\end{equation}

For $0\leq q\leq1$ the channel is CPTP, has $u(\mathcal D_q)=0$, and gives survival probability $(1-q)+q/d$ on every pure input. Thus $R(\mathcal D_q)=\epsilon_S(\mathcal D_q)=q(d-1)/d$. Taking $q=d\varepsilon/(d-1)$ gives $\Phi_S\geq\varepsilon$ for sufficiently small $\varepsilon$ at any $\eta\geq0$.

For the budget-dependent term, take a nonscalar $L\in\mathcal A_S$ and the two-Kraus family of Appendix~\ref{app-robustness-a-physical-family-that-tests-necessity-and-sharpness}. Its image of $I$ is the sum of $K_0K_0^\dagger=I-tL^\dagger L$ and $K_1K_1^\dagger=tLL^\dagger$, giving
\begin{equation}
\label{app-nonunitality-eq-16}
\mathcal E_t(I)-I=t(LL^\dagger-L^\dagger L),
\qquad
u(\mathcal E_t)=\nu_Lt,
\qquad
\nu_L=\frac{\|LL^\dagger-L^\dagger L\|_1}{2d}.
\end{equation}

A normal matrix has orthogonal eigenspaces for distinct eigenvalues. If $L$ were normal, overlapping probes would have equal eigenvalues, and connectivity and spanning would force $L$ to be scalar. Thus $L$ is nonnormal and $\nu_L>0$.

Since $Q_S(L)=0$, the expansions in Eq.~\eqref{app-robustness-eq-15} give $\epsilon_S(\mathcal E_t)\leq B_Lt^2$ and $R(\mathcal E_t)\geq a_Lt$ for small $t$, with fixed positive constants. To satisfy both budgets, keep $t$ within the scales $\sqrt\varepsilon$ and $\eta$. Choose
\begin{equation}
\label{app-nonunitality-eq-17}
t=c_L\min\{\sqrt\varepsilon,\eta\},
\end{equation}

where $B_Lc_L^2\leq1$ and $\nu_Lc_L\leq1$. Since $t\leq c_L\sqrt\varepsilon$, choosing $\varepsilon_0$ small enough ensures the channel is physical and the estimates apply for all $\eta$. Both budgets are satisfied:
\[
\epsilon_S(\mathcal E_t)\leq B_Lc_L^2\varepsilon\leq\varepsilon,
\qquad
u(\mathcal E_t)\leq\nu_Lc_L\eta\leq\eta.
\]
The risk bound gives $\Phi_S\geq a_Lc_L\min\{\sqrt\varepsilon,\eta\}$. Taking the larger of this and the depolarizing bound gives at least half their sum, hence
\[
\Phi_S(\varepsilon,\eta)\geq
\tfrac12\min\{1,a_Lc_L\}\bigl(\varepsilon+\min\{\eta,\sqrt\varepsilon\}\bigr).
\]
If instead $\mu_S>0$, the unrestricted linear robustness theorem and the same depolarizing lower bound give $\Phi_S(\varepsilon,\eta)=\Theta_S(\varepsilon)$, uniformly in $\eta\geq0$ for small $\varepsilon$.

\subsection{A local realization of the qubit witness}
\label{app-nonunitality-a-local-realization-of-the-qubit-witness}

For the equally weighted probes $|0\rangle$ and $|+\rangle$, the commutator gap is $\gamma_{S_2}=1/4$ (Appendix~\ref{app-probe-geometry-computing-the-gaps}). The sharp unital bound is therefore $R\leq4\epsilon_{S_2}/3$.

The nonunital qubit witness from Appendix~\ref{app-probe-geometry-an-exact-qubit-example} has Kraus operators
\begin{equation}
\label{app-nonunitality-eq-18}
K_0=|0\rangle\langle0|+\sqrt{1-2t}|1\rangle\langle1|,
\qquad
K_1=\sqrt{2t}|+\rangle\langle1|,
\qquad0\leq t\leq\frac12.
\end{equation}

Prepare the environment qubit in $|0_e\rangle$ and, when the system is in $|1\rangle$, rotate the environment to $\sqrt{1-2t}|0_e\rangle+\sqrt{2t}|1_e\rangle$. Conditioned on $|1_e\rangle$, a system flip completes amplitude damping with probability $2t$. A Hadamard on the system, conditioned on the same environment state, then sends the jump branch from $|0\rangle$ to $|+\rangle$. This gives the isometry
\begin{equation}
\label{app-nonunitality-eq-19}
\begin{aligned}
|0\rangle&\longmapsto |0\rangle|0_e\rangle,\\
|1\rangle&\longmapsto
\sqrt{1-2t}|1\rangle|0_e\rangle
+\sqrt{2t}|+\rangle|1_e\rangle.
\end{aligned}
\end{equation}

Projecting the environment onto $|0_e\rangle$ and $|1_e\rangle$ gives $K_0$ and $K_1$, respectively. Tracing it out gives the channel. The implementation uses a constant number of local gates and one initialized environment qubit per use.

To compute its nonunitality, write $P_1=|1\rangle\langle1|$ and $P_+=|+\rangle\langle+|$. The two terms in $\mathcal E_t(I)$ are $I-2tP_1$ and $2tP_+$. Subtracting $I$ gives $2t(P_+-P_1)$. Its two eigenvalues are $\pm\sqrt2\,t$, so its trace norm is $2\sqrt2\,t$. Substituting $d=2$ into Eq.~\eqref{app-nonunitality-eq-9} gives
\begin{equation}
\label{app-nonunitality-eq-20}
\mathcal E_t(I)-I
=t\begin{pmatrix}1&1\\1&-1\end{pmatrix},
\qquad
u(\mathcal E_t)=\frac{t}{\sqrt2}.
\end{equation}

On $|+\rangle$, the jump flag has probability $t$, so observing it would reveal a first-order signal. The verifier in Appendices~\ref{app-certification} and~\ref{app-product-channels} receives the reduced system output, whose survival error on $|+\rangle$ is quadratic by Eq.~\eqref{app-probe-geometry-eq-7}, with the cancellation explained in Appendix~\ref{app-robustness-first-order-invisibility-is-a-statement-about-the-loss}. Checking the additional promise $u(\mathcal E)\leq\eta$ can require inputs beyond the menu because $I/d$ may lie outside the convex hull of the probe states.

\section{Adaptive certification and its exact precision coefficient}
\label{app-certification}

This appendix proves Theorem~\ref{thm:certification}. The variance gap gives an upper call bound for a chosen sampling distribution. For the converse, minimax selects error directions for one channel that is hard to detect on every ray. The classical--quantum specialization of the order-$1/2$ amortized channel-divergence method~\cite{wilde-et-al-2020} turns those per-input losses into a lower bound for adaptive verifiers. The final subsections treat trusted reference copies and changes to the acceptance condition.

\subsection{Certification task and allowed inputs}
\label{app-certification-certification-task-and-access-contract}

Fix a finite menu of normalized pure rays $S=\{P_1,\ldots,P_N\}$ in dimension $d\geq2$. The target is a unitary $U$, and the device implements a fixed memoryless channel $\mathcal V$. Test
\begin{equation}
\label{app-certification-eq-1}
H_0:\mathcal V=\mathcal U,
\qquad
H_1:R(\mathcal U^\dagger\circ\mathcal V)\geq\delta,
\end{equation}

where $\mathcal U(X)=UXU^\dagger$ and $R$ is Haar average infidelity. The verifier must keep both the type-I error and the worst-case type-II error at most $\alpha$. We use $0< \alpha\leq\alpha_*< 1/2$ and sufficiently small $\delta>0$. The nonempty alternative range is $0<\delta\leq d/(d+1)$. The upper endpoint follows from Eq.~\eqref{app-robustness-eq-8} and is attained by a traceless unitary such as a cyclic basis shift.

The sample budget $T$ caps device calls on every branch, including protocols that stop early.

\begin{itemize}
\item Before each device call, a common quantum instrument may act on all retained outputs, classical records, and ancillary memory to choose a classical label $i$.
\item The input to that call is a fresh copy of $P_i$, conditionally independent of the retained memory given that classical label.
\item The verifier may keep every output coherently, process outputs and memory jointly, and choose subsequent labels adaptively.
\item A final collective measurement may act on the entire retained state.

\end{itemize}
A prescribed input may be entangled within the register. Each fresh preparation passes through the device once, and the verifier receives the system output. Postselection may use accessible records within the call cap. Error probabilities average over all branches.

For some channel pairs, a sequential strategy that recycles an output as a later input achieves perfect discrimination, and every finite parallel strategy has nonzero error~\cite{harrow-et-al-2010}. Entangling the input with a retained ancilla can improve one-use discrimination for some channel pairs~\cite{sacchi-2005}.

The sharp coefficient applies to the model with ideal target-output projectors $UP_iU^\dagger$. The call cap $T$ counts device uses, while projector implementation contributes to the measurement cost. Appendix~\ref{app-certification-trusted-pure-target-references} gives the call bound with paid pure reference copies. The relative channel $\mathcal E=\mathcal U^\dagger\circ\mathcal V$ expresses the output in target coordinates. A learned device is fixed before collecting fresh data. For a memoryless device, the verifier's stored state carries the dependence on earlier observations.

\subsection{Geometry and risk identities}
\label{app-certification-geometry-and-risk-identities}

For $P_i=|\psi_i\rangle\langle\psi_i|$, let
\begin{equation}
\label{app-certification-eq-2}
e_i(\mathcal E)=1-\Tr\,\bigl[P_i\mathcal E(P_i)\bigr],
\qquad
\epsilon_S(\mathcal E)=\sum_i p_i e_i(\mathcal E),
\end{equation}

and define
\begin{equation}
\label{app-certification-eq-3}
Q_i(A)=\bigl\|(I-P_i)A|\psi_i\rangle\bigr\|^2,
\qquad
Q_p(A)=\sum_i p_iQ_i(A).
\end{equation}

For a fixed positive distribution $p$, denote its gap by $\mu_S$. The menu holds the rays fixed, and $\Delta_N=\{p\in\mathbb R^N:p_i\geq0,\ \sum_i p_i=1\}$ is the set of all sampling distributions on those rays. Optimizing the gap gives
\begin{equation}
\label{app-certification-eq-4}
\mu_*=
\max_{p\in\Delta_N}
\min_{\substack{\Tr\,A=0\\\|A\|_F=1}}Q_p(A).
\end{equation}

The independent sampler draws from the positive support of its distribution. The adaptive verifier can select from the full menu. Sample uniformly with probability $\eta>0$ and from an optimizing distribution otherwise. The resulting distribution is strictly positive and has gap at least $(1-\eta)\mu_*$ by nonnegativity of $Q_i$. This recovers the same supremum as $\eta\to0$.

For Kraus operators $K_a$ of any CPTP channel, the identities from Appendix~\ref{app-robustness-two-exact-kraus-identities} give
\begin{equation}
\label{app-certification-eq-5}
e_i(\mathcal E)=\sum_aQ_i(K_a),
\qquad
R(\mathcal E)
=\frac1{d+1}\sum_a
\left\|K_a-\frac{\Tr\,K_a}{d}I\right\|_F^2.
\end{equation}

Use scalar-shift invariance to apply the gap to the traceless part of each Kraus operator, as in Appendix~\ref{app-robustness-proof-of-the-upper-bound}. With optimizing weights and $\mu_*>0$, summing the resulting inequalities gives
\begin{equation}
\label{app-certification-eq-6}
\epsilon_S(\mathcal E)\geq c\,R(\mathcal E),
\qquad
c=(d+1)\mu_*\in(0,1].
\end{equation}

The absolute gap bound in Appendix~\ref{app-optimal-design-absolute-optimum-and-projective-designs} gives $c\leq1$ on the optimizer's positive support.

\subsection{Adaptive cq fidelity lemma}
\label{app-certification-adaptive-cq-fidelity-lemma}

Use squared Uhlmann fidelity and its square root:
\begin{equation}
\label{app-certification-eq-7}
F(\rho,\sigma)=\|\sqrt\rho\sqrt\sigma\|_1^2,
\qquad
f(\rho,\sigma)=\sqrt{F(\rho,\sigma)}.
\end{equation}

Here $\|X\|_1=\Tr\sqrt{X^\dagger X}$ is the trace norm. The same root-fidelity formula is used for positive subnormalized blocks.

\textbf{Adaptive fidelity lemma.} Let $\rho_i$ and $\sigma_i$ be the two device outputs on input $P_i$, and let $\Omega_0,\Omega_1$ be the two final verifier states before its decision measurement, with all random branches included. Suppose the output states obey
\begin{equation}
\label{app-certification-eq-8}
F(\rho_i,\sigma_i)\geq1-b
\qquad\text{for every }i,
\qquad 0\leq b< 1.
\end{equation}

Every allowed protocol with at most $T$ device calls has final states satisfying
\begin{equation}
\label{app-certification-eq-9}
F(\Omega_0,\Omega_1)\geq(1-b)^T.
\end{equation}

\textbf{Proof.} Let $M_0$ and $M_1$ be the verifier's memory states just before a label choice. The common instrument $\{\mathcal I_i\}$ gives blocks $X_i=\mathcal I_i(M_0)$ and $Y_i=\mathcal I_i(M_1)$. Their traces are the probabilities of choosing label $i$ under the respective hypotheses. Both the branch probabilities and the conditional states can differ.

Retaining the label gives states $\bigoplus_i X_i$ and $\bigoplus_i Y_i$. The square root acts blockwise and the trace norm sums over the blocks, giving root fidelity $\sum_i f(X_i,Y_i)$. The recorded instrument is trace preserving, so monotonicity gives $\sum_i f(X_i,Y_i)\geq f(M_0,M_1)$.

Within branch $i$, the fresh input $P_i$ is a tensor factor independent of the memory. Applying the system channel to the input factor gives blocks $X_i\otimes\rho_i$ and $Y_i\otimes\sigma_i$. The trace norm multiplies across the tensor factors, so
\begin{equation}
\label{app-certification-eq-10}
\begin{aligned}
f\left(\bigoplus_iX_i\otimes\rho_i,
       \bigoplus_iY_i\otimes\sigma_i\right)
&=\sum_i f(X_i,Y_i)f(\rho_i,\sigma_i)\\
&\geq\sqrt{1-b}\sum_i f(X_i,Y_i).
\end{aligned}
\end{equation}

Each round leaves root fidelity at least $\sqrt{1-b}\,f(M_0,M_1)$, and subsequent common processing preserves this lower bound. Starting with identical memories, multiply the per-round factors and square to obtain Eq.~\eqref{app-certification-eq-9}. To pad a stopped branch to $T$ calls, retain its record, send an allowed input on each padding call, and discard those outputs. The verifier decides from the retained state. $\square$

\textbf{Binary-testing converse.} If both errors of a final binary test are at most $\alpha< 1/2$, then
\begin{equation}
\label{app-certification-eq-11}
T\geq
\frac{\log\!\left[1/(4\alpha(1-\alpha))\right]}
     {-\log(1-b)}.
\end{equation}

Write $u$ for the null rejection probability and $v$ for the alternative acceptance probability. In acceptance--rejection order the distributions are $(1-u,u)$ and $(v,1-v)$, with $u,v\leq\alpha$. Their root fidelity is
\[
\sqrt{(1-u)v}+\sqrt{u(1-v)}.
\]
For $u=\sin^2 a$ and $v=\sin^2 a'$ with $a,a'\in[0,\arcsin\sqrt{\alpha}]$, this expression is $\sin(a+a')$. The upper bound $\alpha<1/2$ keeps $a+a'$ below $\pi/2$, so the maximum occurs at $u=v=\alpha$. Squaring gives $4\alpha(1-\alpha)$. By monotonicity, the decision distributions have squared fidelity at least as large as that of the final verifier states. Thus $(1-b)^T\leq 4\alpha(1-\alpha)$, and taking logarithms gives Eq.~\eqref{app-certification-eq-11}. Under perfect completeness $u=0$, the squared binary fidelity is $v\leq\alpha$. This changes the numerator to $\log(1/\alpha)$.

When $b=0$, the two devices give identical outputs on every allowed input. No finite test can then have both errors below $1/2$. The displayed lower bound is interpreted as $+\infty$.

This lemma and converse specialize the cq fidelity framework of Wilde et al.~\cite{wilde-et-al-2020} to the present input model.

\subsection{Fixed-menu minimax duality}
\label{app-certification-fixed-menu-minimax-duality}

Let $\mathcal H_0$ be the complex Hilbert space of traceless $d\times d$ operators with the Hilbert--Schmidt inner product. Write $\mathsf Q_i$ for the positive operator representing $Q_i$ on $\mathcal H_0$. The inner minimization in Eq.~\eqref{app-certification-eq-4} gives the smallest eigenvalue of $\sum_i p_i\mathsf Q_i$. Minimizing its expectation over density operators $Z$ gives the same value: every such expectation is a convex combination of eigenvalues, and a lowest-eigenvector projector attains the minimum. Consequently
\begin{equation}
\label{app-certification-eq-12}
\begin{aligned}
\mu_*
&=\max_{p\in\Delta_N}
   \min_{\substack{Z\succeq0\\\Tr_{\mathcal H_0}Z=1}}
   \sum_i p_i\Tr_{\mathcal H_0}(\mathsf Q_iZ)\\
&=\min_{\substack{Z\succeq0\\\Tr_{\mathcal H_0}Z=1}}
   \max_i\Tr_{\mathcal H_0}(\mathsf Q_iZ).
\end{aligned}
\end{equation}

The payoff is bilinear on compact convex domains, so convex minimax exchanges the optimizations and both optima are attained. The inner maximization then puts all weight on a largest term, giving the maximum over $i$.

Choose a minimizing $Z$ and factor it as
\begin{equation}
\label{app-certification-eq-13}
Z=\sum_a|L_a\rangle\!\rangle\langle\!\langle L_a|,
\qquad
\Tr\,L_a=0,
\qquad
\sum_a\|L_a\|_F^2=1.
\end{equation}

For each nonzero eigenvalue $z_a$ of $Z$ with normalized eigenvector $|A_a\rangle\!\rangle$, take $L_a=\sqrt{z_a}\,A_a$. This gives at most $d^2-1$ traceless terms with squared Frobenius norms summing to $\Tr Z=1$. The expectation of $\mathsf Q_i$ in that mixture gives the simultaneous constraint
\begin{equation}
\label{app-certification-eq-14}
\sum_aQ_i(L_a)
=\Tr_{\mathcal H_0}(\mathsf Q_iZ)
\leq\mu_*.
\end{equation}

One minimizing $Z$ gives this bound on every ray, including rays chosen adaptively.

\subsection{Exact multi-Kraus adversary}
\label{app-certification-exact-multi-kraus-adversary}

Set
\begin{equation}
\label{app-certification-eq-15}
B=\sum_aL_a^\dagger L_a,
\qquad
K_a(t)=\sqrt t\,L_a,
\qquad
K_0(t)=\sqrt{I-tB},
\qquad 0\leq t\leq1.
\end{equation}

The operator $B$ is positive and has trace $\sum_a\|L_a\|_F^2=1$, so each eigenvalue is at most one and $B\preceq I$. These Kraus operators define an exact CPTP channel because
\begin{equation}
\label{app-certification-eq-16}
K_0(t)^\dagger K_0(t)+\sum_aK_a(t)^\dagger K_a(t)
=I-tB+tB=I.
\end{equation}

With an orthonormal environment flag $|a\rangle_{\rm env}$ for each Kraus operator, the dilation maps the input to
\[
V_t|\psi\rangle=\sum_{a\geq0}K_a(t)|\psi\rangle\otimes|a\rangle_{\rm env}.
\]
Equation~\eqref{app-certification-eq-16} makes $V_t$ an isometry. Discarding the flag realizes the mixture of error directions as one fixed system channel $\mathcal E_t$ on every call.

The traceless $a\geq1$ terms contribute $t\sum_{a\geq1}\|L_a\|_F^2=t$ to the squared norms in Eq.~\eqref{app-certification-eq-5}. The remaining term gives
\begin{equation}
\label{app-certification-eq-17}
R(\mathcal E_t)
=\frac{t}{d+1}
 +\frac1{d+1}
  \left\|K_0(t)-\frac{\Tr\,K_0(t)}{d}I\right\|_F^2
\geq\frac{t}{d+1}.
\end{equation}

Each training error obeys
\begin{equation}
\label{app-certification-eq-18}
e_i(\mathcal E_t)
=t\sum_aQ_i(L_a)+Q_i(K_0(t))
\leq t\mu_*+Q_i(K_0(t)).
\end{equation}

Scalar-shift invariance of $Q_i$ and spectral calculus give a uniform quadratic bound on the residual
\begin{equation}
\label{app-certification-eq-19}
\begin{aligned}
Q_i(K_0(t))
&=Q_i(K_0(t)-I)\\
&\leq\|K_0(t)-I\|_F^2\\
&\leq\frac{t^2\Tr\,B^2}
             {(1+\sqrt{1-t})^2}\\
&\leq t^2.
\end{aligned}
\end{equation}

Since $I-P_i$ is a projection and $\psi_i$ is normalized, $Q_i(M)\leq\|M\|_F^2$. To bound $\|K_0(t)-I\|_F^2$, sum the squared eigenvalue differences of $K_0(t)$ from $I$. Each eigenvalue $0\leq\lambda\leq1$ of $B$ contributes
\[
\bigl(1-\sqrt{1-t\lambda}\bigr)^2
=\frac{t^2\lambda^2}{(1+\sqrt{1-t\lambda})^2}
\leq \frac{t^2\lambda^2}{(1+\sqrt{1-t})^2}.
\]
The sum is bounded using $\sum_\lambda\lambda^2=\Tr\,B^2\leq(\Tr\,B)^2=1$.

Thus one physical channel simultaneously satisfies
\begin{equation}
\label{app-certification-eq-20}
R(\mathcal E_t)\geq\frac{t}{d+1},
\qquad
\max_i e_i(\mathcal E_t)\leq t\mu_*+t^2.
\end{equation}

For the device $\mathcal V_t=\mathcal U\circ\mathcal E_t$, unitary invariance reduces the output fidelity to $F(P_i,\mathcal E_t(P_i))$. For a rank-one projector $P$, the squared trace norm of $P\sqrt\rho$ is the sole possible nonzero eigenvalue of $P\rho P$, namely $\Tr(P\rho)$. Thus the squared fidelity on label $i$ is $1-e_i(\mathcal E_t)$.

\subsection{Finite-sample sandwich}
\label{app-certification-finite-sample-sandwich}

Assume $\mu_*>0$ and define
\begin{equation}
\label{app-certification-eq-21}
c=(d+1)\mu_*,
\qquad
b_\delta=c\delta+(d+1)^2\delta^2.
\end{equation}

Let $T_{\rm opt}(\delta,\alpha)$ be the smallest worst-case call cap among all allowed verifiers. Whenever $0< \delta\leq1/(d+1)$ and $b_\delta< 1$,
\begin{equation}
\label{app-certification-eq-22}
\boxed{
\frac{\log\!\left[1/(4\alpha(1-\alpha))\right]}
     {-\log(1-b_\delta)}
\ \leq\ T_{\rm opt}(\delta,\alpha)
\ \leq\
\left\lceil
\frac{\log(1/\alpha)}{-\log(1-c\delta)}
\right\rceil.
}
\end{equation}

For the lower bound, choose $t=(d+1)\delta\leq1$ in Eq.~\eqref{app-certification-eq-20}. The alternative has risk at least $\delta$ and squared fidelity at least $1-b_\delta$ with the target on every ray. Equation~\eqref{app-certification-eq-11} applies to this alternative for every adaptive verifier.

For the upper bound, draw labels independently from the optimizing distribution and test each output immediately. The trials are independent because the inputs are fresh and the device is memoryless. Each rejects with probability $\epsilon_S(\mathcal E)$, at least $c\delta$ under every alternative by Eq.~\eqref{app-certification-eq-6}. Accepting only if all trials pass gives false-acceptance probability at most $(1-c\delta)^T$ and perfect completeness. Solving for error at most $\alpha$ gives the ceiling in Eq.~\eqref{app-certification-eq-22}.

Retaining the spectral denominator in Eq.~\eqref{app-certification-eq-19} sharpens the finite lower bound without changing the limit.

\subsection{Completion of the proof of Theorem~\ref{thm:certification}}
\label{app-certification-exact-optimized-sampling-coefficient}

The finite-sample bounds give the joint limit in Eq.~\eqref{main-certification-coefficient} by separating the risk and confidence terms. The logarithmic denominators have the expansions
\begin{equation}
\label{app-certification-eq-24}
-\log(1-b_\delta)=c\delta+O_{d,S}(\delta^2),
\qquad
-\log(1-c\delta)=c\delta+O_{d,S}(\delta^2),
\end{equation}
and the confidence factors satisfy
\begin{equation}
\label{app-certification-eq-25}
\frac{\log[1/(4\alpha(1-\alpha))]}{\log(1/\alpha)}\longrightarrow1.
\end{equation}

With $L_\alpha=\log[1/(4\alpha(1-\alpha))]$, the normalized lower bound is
\[
\frac{\delta}{-\log(1-b_\delta)}
\frac{L_\alpha}{\log(1/\alpha)}.
\]
The first factor tends to $1/c$ as $\delta\to0$. Since $L_\alpha=\log(1/\alpha)-\log[4(1-\alpha)]$, the second tends to one as $\alpha\to0$. The upper bound consists of the risk factor, with $c\delta$ in place of $b_\delta$, and a ceiling contribution at most $\delta/\log(1/\alpha)$. The risk and confidence parameters can therefore tend to zero independently.

At fixed confidence $\alpha<1/2$, the two logarithmic numerators retain their distinct values, giving
\begin{equation}
\label{app-certification-eq-26}
\frac{\log[1/(4\alpha(1-\alpha))]}{c}
\leq\liminf_{\delta\to0}\delta T_{\rm opt}(\delta,\alpha)
\leq\limsup_{\delta\to0}\delta T_{\rm opt}(\delta,\alpha)
\leq\frac{\log(1/\alpha)}{c}.
\end{equation}

Let $T_{\rm pc}(\delta,\alpha)$ be the optimal call cap with zero type-I error and worst-case type-II error at most $\alpha$. Perfect completeness makes the converse numerator $\log(1/\alpha)$, and the all-pass protocol has perfect completeness. The common confidence factor cancels, giving
\begin{equation}
\label{app-certification-eq-27}
\lim_{\delta\to0}
\frac{\delta\,T_{\rm pc}(\delta,\alpha)}{\log(1/\alpha)}
=\frac1c.
\end{equation}

All these coefficient statements keep $d$ and $S$ fixed. In $b_\delta$, the ratio of the quadratic term to the linear term is $(d+1)\delta/\mu_*$. Thus $\delta\ll\mu_* /(d+1)$ makes that contribution negligible. Appendix~\ref{app-product-channels} gives repetition bounds uniform in register size for product channels.

The ratio $L_\alpha/\log(1/\alpha)$ extends continuously to one at $\alpha=0$ and is positive for $\alpha<1/2$. Its minimum on $[0,\alpha_*]$ is positive for a fixed $\alpha_*<1/2$. The lower bounds can therefore use $\log(1/\alpha)$ uniformly over $0<\alpha\leq\alpha_*$.

Combining Theorems~\ref{thm:certification} and~\ref{thm:design} (Appendix~\ref{app-optimal-design}) gives a lower bound for every allowed strategy. For a menu with $\mu_*>0$ and $d< N\leq d^2$,
\begin{equation}
\label{app-certification-eq-28}
\lim_{\substack{\delta\to0\\\alpha\to0}}
\frac{\delta\,T_{\rm opt}(\delta,\alpha)}{\log(1/\alpha)}
\geq\frac{d(N-1)}{(d+1)(N-d)}.
\end{equation}

An equiangular-tight-frame menu attains this bound when it exists. Weighted projective $2$-designs attain the absolute lower coefficient $1$.

\subsection{Complete precision trichotomy}
\label{app-certification-complete-precision-trichotomy}

Appendix~\ref{app-robustness-exact-identification} characterizes exact identification by spanning and overlap connectivity. The gap criterion in Appendix~\ref{app-robustness-the-linear-robustness-theorem} uses the common right-eigenoperator space
\begin{equation}
\label{app-certification-eq-29}
\mathcal A_S=
\{L:L|\psi_i\rangle\in\mathbb C|\psi_i\rangle
\text{ for every }i\}.
\end{equation}

Fix $d,S$ and a confidence ceiling $0<\alpha_*<1/2$. Uniformly for $0<\alpha\leq\alpha_*$, the optimal precision orders are

\begin{center}
\small
\setlength{\tabcolsep}{4pt}
\begin{tabular}{@{}p{0.43\linewidth}p{0.52\linewidth}@{}}
\toprule
Geometry of the available rays & Optimal sample order as $\delta\to0$ \\
\midrule
Rays span, overlap graph is connected, and $\mathcal A_S=\mathbb CI$ & $\Theta_{d,S}(\delta^{-1}\log(1/\alpha))$ \\
Rays span, overlap graph is connected, and $\mathcal A_S\neq\mathbb CI$ & $\Theta_{d,S}(\delta^{-2}\log(1/\alpha))$ \\
Rays do not span, or the overlap graph is disconnected & Impossible below a positive fixed risk threshold \\
\bottomrule
\end{tabular}
\end{center}

For the first row the scalar kernel gives a positive gap. The finite-sample denominators scale as $\delta$, and the confidence comparison above makes the order uniform up to $\alpha_*$.

For the second row, take a traceless $L\in\mathcal A_S$ with $\|L\|_F=1$ and use the witness from Appendix~\ref{app-robustness-a-physical-family-that-tests-necessity-and-sharpness}. Its two Kraus operators for $0\leq t\leq1$ are
\begin{equation}
\label{app-certification-eq-30}
K_1=\sqrt t\,L,
\qquad K_0=\sqrt{I-tL^\dagger L}
\end{equation}

The channel has $R\geq t/(d+1)$ by Eq.~\eqref{app-certification-eq-5}. Since $Q_i(L)=0$ on every ray, Eq.~\eqref{app-certification-eq-19} bounds each $e_i=Q_i(K_0)$ by $O_{d,S}(t^2)$. The connected-spanning argument for the positive weighted quadratic coefficient $Q_p(L^\dagger L)/4$ is in Appendix~\ref{app-robustness-the-sharp-robustness-trichotomy}. For certification the bound needs to hold on each ray: set $t=(d+1)\delta$ and use the simultaneous $O_{d,S}(\delta^2)$ losses in Eq.~\eqref{app-certification-eq-11}. This proves the $\Omega(\delta^{-2}\log(1/\alpha))$ lower bound for adaptive collective protocols.

For the matching upper bound, fix any positive weights and define
\begin{equation}
\label{app-certification-eq-31}
m_S(\delta)
=\min_{\substack{\mathcal E\ \text{CPTP}\\R(\mathcal E)\geq\delta}}
  \epsilon_S(\mathcal E).
\end{equation}

The continuous loss attains its minimum because the admissible Choi matrices form a nonempty closed subset of the positive semidefinite set of fixed trace. The square-root bound from Mayer--Knill~\cite{mayer-knill-2018}, converted in Appendix~\ref{app-robustness-imported-square-root-upper-bound}, gives a finite $B_S$ such that
\begin{equation}
\label{app-certification-eq-32}
R(\mathcal E)\leq B_S\sqrt{\epsilon_S(\mathcal E)}.
\end{equation}

Since $-\log(1-m_S(\delta))\geq m_S(\delta)\geq\delta^2/B_S^2$, the all-pass argument gives
\begin{equation}
\label{app-certification-eq-33}
T\leq
\left\lceil
\frac{\log(1/\alpha)}{-\log(1-m_S(\delta))}
\right\rceil
=O_{d,S}\bigl(\delta^{-2}\log(1/\alpha)\bigr).
\end{equation}

For the third row, use a phase unitary $W$ from Appendix~\ref{app-robustness-exact-identification}. It has positive risk $r_W$ and identical outputs to the identity on all allowed inputs. Every adaptive protocol then has the same decision distribution under the two hypotheses. For $0<\delta\leq r_W$, the two errors sum to one.

The constants in the first two rows can depend on the menu's conditioning, its fixed positive weights, dimension, and the fixed confidence ceiling $\alpha_*$. The classification describes precision dependence at fixed dimension and menu.

\subsection{Trusted pure target references}
\label{app-certification-trusted-pure-target-references}

For each label $i$, compare the output $\sigma_i=\mathcal V(P_i)$ with a fresh trusted pure reference $\tau_i=UP_iU^\dagger$. Let $\mathbb F$ exchange the two registers. The antisymmetric projector is $(I-\mathbb F)/2$, and expanding the trace in a product basis gives
\[
\Tr[\mathbb F(A\otimes B)]
=\sum_{j,k}\langle k|A|j\rangle\,\langle j|B|k\rangle
=\Tr(AB).
\] The probability of the antisymmetric outcome is
\begin{equation}
\label{app-certification-eq-34}
\Tr\,\left[
\frac{I-\mathbb F}{2}(\sigma_i\otimes\tau_i)
\right]
=\frac{1-\Tr(\sigma_i\tau_i)}2
=\frac{e_i}{2}.
\end{equation}

This is the standard state-comparison identity of Barnett, Chefles, and Jex~\cite{barnett-et-al-2003}, extended to the mixed device output by linearity. Exact equality of the two pure outputs under the null gives perfect completeness.

In the positive-gap regime, independent sampling of the optimal labels gives each alternative average rejection probability $\epsilon_S(\mathcal E)/2\geq c\delta/2$. The all-pass cap therefore changes to
\begin{equation}
\label{app-certification-eq-35}
T_{\rm SWAP}\leq
\left\lceil
\frac{\log(1/\alpha)}{-\log(1-c\delta/2)}
\right\rceil.
\end{equation}

Here $T_{\rm SWAP}$ is a sufficient device-call cap. Each trial also uses one target-output copy and incurs its preparation cost. For $n$ output qubits, a controlled-SWAP test~\cite{buhrman-et-al-2001} uses order $n$ local controlled swaps and one control qubit. In the singular regime, replacing $c\delta$ by $m_S(\delta)$ gives the order-$\delta^{-2}$ bound.

In label block $i$, the common reference contributes the factor $f(\tau_i,\tau_i)=1$, preserving the adaptive lower bound. The SWAP protocol therefore attains the same precision orders. The denominator in Eq.~\eqref{app-certification-eq-35} is asymptotic to $c\delta/2$, so doubling the leading ideal-projector call bound suffices in the positive-gap regime.

\subsection{Shrinking acceptance tolerance}
\label{app-certification-shrinking-acceptance-tolerance}

Suppose a randomized survival test has rejection probability $q=\epsilon_S$ and the hypotheses guarantee
\begin{equation}
\label{app-certification-eq-36}
H_{\rm good}:q\leq\theta m,
\qquad
H_{\rm bad}:q\geq m,
\qquad 0\leq\theta< 1,
\qquad 0< m\leq1.
\end{equation}

Reject if the frequency in independent trials exceeds $(1+\theta)m/2$. Coupling Bernoulli trials by common uniform draws shows that the frequency increases stochastically with $q$. The largest errors occur at the endpoints $\theta m$ and $m$, both at distance $h=(1-\theta)m/2$ from the threshold. A multiplicative Chernoff bound~\cite{chernoff-1952,boucheron-et-al-2013} for a relative downward deviation $h/m$ at $q=m$ gives error at most
\begin{equation}
\label{app-certification-eq-37}
\exp\left[-\frac{T(1-\theta)^2m}{8}\right]
\end{equation}

for the bad class. At the good endpoint $q=\theta m$, rejection requires an upward deviation by $h$. A Bernstein upper-tail bound~\cite{boucheron-et-al-2013} has exponent at least $Th^2/[2(\theta m+h/3)]$, using variance at most $\theta m$ and range bounded by one. Since $\theta m+h/3\leq m$, this gives the same sufficient exponent. Therefore the choice
\begin{equation}
\label{app-certification-eq-38}
T=
\left\lceil
\frac{8}{(1-\theta)^2m}\log\frac1\alpha
\right\rceil
\end{equation}

controls both errors.

For fixed $\theta$, use $m=c\delta$ in the linear regime or the bound of order $\delta^2$ from Eq.~\eqref{app-certification-eq-32} in the singular regime. The precision orders match the original task because the good class includes the exact target and the bad class includes every channel of risk at least $\delta$.

\subsection{Fixed noisy baselines have a different exponent}
\label{app-certification-fixed-noisy-baselines-have-a-different-exponent}

Consider the depolarizing channel whose survival infidelity on every pure input is $r$:
\begin{equation}
\label{app-certification-eq-39}
\mathcal D_r(X)
=\left(1-\frac{dr}{d-1}\right)X
 +\frac{r}{d-1}\Tr(X)I,
\qquad 0\leq r\leq\frac{d-1}{d}.
\end{equation}

The map mixes the identity and complete depolarization with depolarizing weight $dr/(d-1)\in[0,1]$, so it is CPTP. Its output on a pure input is
\begin{equation}
\label{app-certification-eq-40}
\mathcal D_r(P)
=(1-r)P+\frac{r}{d-1}(I-P).
\end{equation}

Fix $0<r_0<(d-1)/d$ and take $\delta>0$ small enough to keep $\mathcal D_{r_0+\delta}$ physical. The Haar risks are $r_0$ and $r_0+\delta$ because the loss is constant over pure inputs. The two outputs commute, with one eigenvalue on the input ray and the remaining weight spread uniformly over $d-1$ eigenvectors. In the root fidelity, the sum over the complement cancels this multiplicity, giving
\begin{equation}
\label{app-certification-eq-41}
\begin{aligned}
F
&=\left[
\sqrt{(1-r_0)(1-r_0-\delta)}
+\sqrt{r_0(r_0+\delta)}
\right]^2\\
&=1-\frac{\delta^2}{4r_0(1-r_0)}+O_{r_0}(\delta^3).
\end{aligned}
\end{equation}

To expand the root fidelity, use $(1-r_0)\sqrt{1-\delta/(1-r_0)}$ and $r_0\sqrt{1+\delta/r_0}$. The linear terms $\mp\delta/2$ cancel. The second-order sum before squaring is $-\delta^2/[8r_0(1-r_0)]$.

Here $b=1-F$ has the same value on every ray. For small $\delta$, the denominator in Eq.~\eqref{app-certification-eq-11} is $-\log F=b+O(b^2)$ and is of order $\delta^2/[r_0(1-r_0)]$. With the fixed confidence ceiling, the classical--quantum converse therefore imposes
\begin{equation}
\label{app-certification-eq-42}
T=\Omega\left(
r_0(1-r_0)\,\delta^{-2}\log\frac1\alpha
\right)
\end{equation}

for any menu under the same fresh-input access contract. An estimator accurate to less than $\delta/2$ with probability at least $1-\alpha$ under each channel would yield this test by thresholding at $r_0+\delta/2$. The bound applies to risk estimation at that precision, even with an optimal projective design.

\section{Uniform bounds for independent qubit channels}
\label{app-product-channels}

The proof of Theorem~\ref{thm:product} starts from product identities that express full-register risk and probe loss in terms of local errors. For two directions, they give the square-root upper bound and calibrate the hard channel at a fixed global risk. For three probes, a product inequality transfers the local linear bound to the register, and noise at a single site gives the sharp coefficient.

\subsection{Independent-channel model and task}
\label{app-product-channels-product-promise-and-operational-task}

Let $n\ge1$, set $D=2^n$, and suppose the relative channel is
\begin{equation}
\label{app-product-channels-eq-1}
\mathcal E=\bigotimes_{j=1}^n\mathcal E_j,
\end{equation}

where each $\mathcal E_j$ is an arbitrary qubit CPTP map fixed across uses. The maps may vary by site. The target is a product unitary. The full-channel factorization gives the product formulas for probe survival and entanglement fidelity below.

For one global use, the verifier prepares an $n$-qubit product input and receives all $n$ output qubits. The verifier has the fresh-input access specified in Appendix~\ref{app-certification}, including arbitrary quantum output memory and adaptive classical choice of later inputs. Its task is to distinguish the exact target from $R(\mathcal E)\ge\delta$, with both errors at most $\alpha$. We write $T_k^\star(\delta,\alpha)$ for the smallest worst-case global call cap on the menu $S_k$. The upper bounds use ideal target-output tests on the global menus
\begin{equation}
\label{app-product-channels-eq-2}
S_2=\{|0\rangle^{\otimes n},|+\rangle^{\otimes n}\},
\qquad
S_3=S_2\cup\{|1\rangle^{\otimes n}\},
\quad |+\rangle=\frac{|0\rangle+|1\rangle}{\sqrt2}.
\end{equation}

The two-direction lower bound includes all $2^n$ mixed tensor strings of local $|0\rangle$ and $|+\rangle$ inputs.

\subsection{Product identities}
\label{app-product-channels-product-identities}

For a probe label $r$, let $P_r=\ket r\bra r$. The local entanglement infidelity enters the full-register risk, and the local survival loss enters the corresponding probe test. Write
\begin{equation}
\label{app-product-channels-eq-3}
x_j=1-F_e(\mathcal E_j),\qquad
y_{jr}=1-\Tr[P_r\mathcal E_j(P_r)],
\qquad r\in\{0,+,1\}.
\end{equation}

For a vector with coordinates in $[0,1]$, define
\begin{equation}
\label{app-product-channels-eq-4}
F(z)=1-\prod_{j=1}^n(1-z_j).
\end{equation}

The quantity $F(z)$ is the probability of at least one failure in independent trials with failure probabilities $z_j$. In particular, it is nondecreasing in every coordinate, and $z_j\le F(z)\le\sum_k z_k$ for each $j$. For a uniform product probe $P_r^{\otimes n}$, the output and the ideal relative-output projector both factor, so the survival probability is
\[
 \Tr\!\left[P_r^{\otimes n}\mathcal E(P_r^{\otimes n})\right]
 =\prod_{j=1}^n\Tr[P_r\mathcal E_j(P_r)]
 =\prod_{j=1}^n(1-y_{jr}).
\]
The same argument for a mixed tensor string uses the direction chosen at each site.

To express the risk, choose local Kraus operators $K_{ja}$ for $\mathcal E_j$ and use the tensor products $\bigotimes_j K_{ja_j}$ as Kraus operators for the product channel. Its entanglement fidelity factors because traces multiply under tensor products and the sum over Kraus labels $a_1,\ldots,a_n$ separates. In the normalization of Appendix~\ref{app-robustness-two-exact-kraus-identities}, this gives
\[
 F_e(\mathcal E)=\prod_{j=1}^n\left(\frac14\sum_a|\Tr K_{ja}|^2\right)
 =\prod_{j=1}^n(1-x_j).
\]
Writing $x=(x_1,\ldots,x_n)$ and $y_r=(y_{1r},\ldots,y_{nr})$ therefore gives
\begin{equation}
\label{app-product-channels-eq-5}
R(\mathcal E)=\frac D{D+1}F(x),
\qquad
Y_r:=e_{r^{\otimes n}}(\mathcal E)=F(y_r).
\end{equation}

The risk averages over all pure inputs of the $D$-dimensional register, including entangled states.

If $Y_r=0$, the product of the local survival probabilities equals one, forcing each local probability to equal one. Thus exact survival of both global probes in $S_2$ makes each map preserve $|0\rangle$ and $|+\rangle$. The rays are spanning and nonorthogonal, so they identify the local identity channel (Appendix~\ref{app-robustness-exact-identification}). Hence $S_2$ identifies the product target.

\subsection{Matching two-direction law}
\label{app-product-channels-matching-two-direction-law}

For any fixed $0<\alpha_*< 1/2$, there are positive constants independent of $n,\delta,\alpha$ such that, for $0< \delta\le1/8$ and $0< \alpha\le\alpha_*$,
\begin{equation}
\label{app-product-channels-eq-6}
T_2^\star(\delta,\alpha)
=\Theta\!\left(\frac n{\delta^2}\log\frac1\alpha\right).
\end{equation}

The same order holds if the two-input menu is expanded to all tensor strings from the two local directions.

\subsubsection{Upper bound}
\label{app-product-channels-upper-bound}

The fixed two-ray qubit robustness theorem gives an absolute constant $C_0$ such that
\begin{equation}
\label{app-product-channels-eq-7}
x_j\le C_0\sqrt{y_{j0}+y_{j+}}.
\end{equation}

The square-root bound in Appendix~\ref{app-robustness-imported-square-root-upper-bound} applies with local mean loss $(y_{j0}+y_{j+})/2$ and Haar risk $2x_j/3$. Since the two rays and the qubit dimension are fixed, $C_0$ is independent of $n$ and of the local channels.

Let $\epsilon_2=(Y_0+Y_+)/2$. If $\epsilon_2\le1/4$, then each $Y_r\le1/2$ and
\begin{equation}
\label{app-product-channels-eq-8}
\sum_jy_{jr}
\le -\sum_j\log(1-y_{jr})
=-\log(1-Y_r)
\le2Y_r.
\end{equation}

The first inequality uses $-\log(1-y)\ge y$, and the last uses $-\log(1-Y_r)\le Y_r/(1-Y_r)\le2Y_r$. Adding the two inequalities for $r=0,+$ bounds $\sum_j(y_{j0}+y_{j+})$ by $4\epsilon_2$. Cauchy--Schwarz bounds the sum of the sitewise square roots by the square root of the product of $n$ and this sum. Since $F(x)\le\sum_jx_j$ and $D/(D+1)\le1$, we obtain
\begin{equation}
\label{app-product-channels-eq-9}
R(\mathcal E)
\le C_0\sum_j\sqrt{y_{j0}+y_{j+}}
\le2C_0\sqrt{n\epsilon_2}.
\end{equation}

For $\epsilon_2>1/4$, we have $\sqrt{n\epsilon_2}>1/2$, so $R\le1$ gives the same inequality with an enlarged absolute constant $C$. Every alternative with $R\ge\delta$ then has $\epsilon_2\ge\delta^2/(C^2n)$. The independent all-pass test of Appendix~\ref{app-certification-finite-sample-sandwich}, with equal weights on the two global probes, has false-acceptance probability at most $\exp[-T\delta^2/(C^2n)]$ and perfect completeness. The protocol gives the same upper bound on the full string menu by using those two global probes.

\subsubsection{Lower bound and its uniform calibration}
\label{app-product-channels-lower-bound-and-its-uniform-calibration}

Use the trace-preserving qubit witness of Appendix~\ref{app-probe-geometry-an-exact-qubit-example} at every site. In the notation here its Kraus operators are
\begin{equation}
\label{app-product-channels-eq-10}
K_0(t)=\begin{pmatrix}1&0\\0&s\end{pmatrix},
\qquad
K_1(t)=\sqrt t\begin{pmatrix}0&1\\0&1\end{pmatrix},
\qquad s=\sqrt{1-2t}.
\end{equation}

On the physical interval $0\le t\le1/2$, Eq.~\eqref{app-probe-geometry-eq-7} gives $y_0$ and $y_+$. Multiplying the local Haar risk in Eq.~\eqref{app-probe-geometry-eq-8} by $3/2$ gives the entanglement infidelity $x(t)$. Thus
\begin{equation}
\label{app-product-channels-eq-11}
y_0=0,\qquad
y_+=\frac{(1-s)^2}{4}=\frac{t^2}{(1+s)^2},
\qquad
x(t)=\frac{1-s}{2}+\frac t4.
\end{equation}

With this channel on all $n$ sites, Eq.~\eqref{app-product-channels-eq-5} gives $R=D[1-(1-x(t))^n]/(D+1)$. The equality $R=\delta$ fixes the local channel strength through
\begin{equation}
\label{app-product-channels-eq-12}
x(t)=x_\delta
:=1-\left(1-\frac{D+1}{D}\delta\right)^{1/n}.
\end{equation}

For $0< \delta\le1/8$, put $z=(D+1)\delta/D$. Since $D\ge2$, we have $\delta\le z\le3\delta/2\le3/16$. For the lower bound, $z=1-(1-x_\delta)^n\le nx_\delta$. For the upper bound, write $x_\delta=1-\exp[\log(1-z)/n]$ and use
\[
 x_\delta\le\frac{-\log(1-z)}{n}
 \le\frac{z}{n(1-z)}
 \le\frac{24}{13}\frac{\delta}{n}.
\]
Also, $x_\delta\le z$ since taking an $n$th root of a number in $[0,1]$ cannot decrease it. These estimates give
\begin{equation}
\label{app-product-channels-eq-13}
\frac\delta n\le x_\delta\le\frac{2\delta}{n},
\qquad x_\delta\le\frac3{16}.
\end{equation}

The function $x(t)$ is continuous and strictly increasing, with $x(0)=0$ and $x(1/4)=9/16-1/(2\sqrt2)>3/16$. Thus a unique solution of Eq.~\eqref{app-product-channels-eq-12} lies in $[0,1/4]$. In this interval, $1/2\le s\le1$ and $x(t)=t/(1+s)+t/4$. It follows that $3t/4\le x(t)\le t$. Together with Eq.~\eqref{app-product-channels-eq-13} this gives
\begin{equation}
\label{app-product-channels-eq-14}
\frac\delta n\le t\le\frac{8\delta}{3n}.
\end{equation}

A string with $k$ occurrences of $|+\rangle$ has failure probability $1-(1-y_+)^k$. The all-plus string maximizes this probability because the remaining inputs have $y_0=0$. Since $s\ge1/2$, the formula for $y_+$ gives $y_+\le t^2/2$. The probability of any site failing is at most the sum of the site failure probabilities. Thus for every allowed string the loss is at most
\begin{equation}
\label{app-product-channels-eq-15}
b=1-(1-y_+)^n
\le ny_+
\le\frac{nt^2}{2}
\le\frac{32\delta^2}{9n}.
\end{equation}

The ideal output is pure, so its squared fidelity with the alternative output is at least $1-b$ on every allowed string. The adaptive binary-testing bound of Appendix~\ref{app-certification-adaptive-cq-fidelity-lemma} gives
\begin{equation}
\label{app-product-channels-eq-16}
T\ge
\frac{\log[1/(4\alpha(1-\alpha))]}{-\log(1-b)}
=\Omega\!\left(\frac n{\delta^2}\log\frac1\alpha\right).
\end{equation}

Equation~\eqref{app-product-channels-eq-15} gives $b\le1/(18n)\le1/18$, so the denominator is at most $2b$. The uniform confidence comparison in Appendix~\ref{app-certification-exact-optimized-sampling-coefficient} bounds the numerator below by a positive multiple of $\log(1/\alpha)$ for $0<\alpha\le\alpha_*$. This gives a lower bound uniform in $n,\delta,\alpha$. The same alternative works for every input label, so the bound covers adaptive choices and collective measurements of retained outputs.

\subsection{An exact three-probe inequality}
\label{app-product-channels-an-exact-three-probe-inequality}

The sharp uniform qubit coefficient in Eq.~\eqref{app-probe-geometry-eq-11} is $2+\sqrt2$. Apply it to mean loss $\sum_r y_{jr}/3$ and local Haar risk $2x_j/3$ to get
\begin{equation}
\label{app-product-channels-eq-17}
x_j\le\kappa\sum_{r\in\{0,+,1\}}y_{jr},
\qquad \kappa=\frac{2+\sqrt2}{2}\ge1.
\end{equation}

The following inequality transfers the bound from local losses to the observed global losses $F(y_r)$ with no factor of $n$.

\subsubsection{Product inequality}
\label{app-product-channels-product-inequality}

If $x_j\le\sum_r c_r y_{jr}$ and every $c_r\ge1$, then
\begin{equation}
\label{app-product-channels-eq-18}
F(x)\le\sum_r c_r F(y_r).
\end{equation}

Let $w=\sum_r c_r y_r$. If some $w_j\ge1$, then $\sum_r c_rF(y_r)\ge w_j\ge1$ since $F(y_r)\ge y_{jr}$. This proves the bound because $F(x)\le1$. If every $w_j<1$, then $w$ and each $c_r y_r$ belong to the unit cube. Since $x_j\le w_j$ for each $j$, monotonicity gives the first inequality in the chain
\begin{equation}
\label{app-product-channels-eq-19}
F(x)\le F\!\left(\sum_r c_ry_r\right)
\le\sum_r F(c_ry_r)
\le\sum_r c_rF(y_r).
\end{equation}

For the second inequality, put $z_r=c_r y_r$. Independently at each site $j$, select one category $r$ with probability $(z_r)_j$ or no category with the remaining probability. Category $r$ occurs at some site with probability $F(z_r)$, and at least one category occurs with probability $F(\sum_r z_r)$. A union bound over categories gives the inequality.

For the third inequality, fix a nonnegative vector $z$ and restrict $t$ to the interval where $tz$ lies in the cube. The function $f(t)=F(tz)$ has $f(0)=0$ and
\begin{equation}
\label{app-product-channels-eq-20}
f''(t)=-\sum_{i\ne j}z_i z_j\prod_{k\ne i,j}(1-tz_k)\le0
\end{equation}

throughout that interval. If $c\ge1$ and $cz$ lies in the cube, concavity at $1=(1/c)c+(1-1/c)0$ gives $f(1)\ge f(c)/c$. Hence $F(cz)\le cF(z)$, which proves the last inequality in Eq.~\eqref{app-product-channels-eq-19}.

\subsubsection{Global coefficient}
\label{app-product-channels-global-coefficient}

With uniform three-probe loss $\epsilon_3=(Y_0+Y_++Y_1)/3$, apply Eq.~\eqref{app-product-channels-eq-18} with $c_r=\kappa$. It gives $F(x)\le\kappa\sum_r Y_r=3\kappa\epsilon_3$. The Haar normalization in Eq.~\eqref{app-product-channels-eq-5} gives the exact, global bound
\begin{equation}
\label{app-product-channels-eq-21}
R(\mathcal E)
\le\frac{3(2+\sqrt2)D}{2(D+1)}\epsilon_3.
\end{equation}

With noise at site $j$ and identity channels elsewhere, $F(x)=x_j$ and $Y_r=y_{jr}$. Equation~\eqref{app-product-channels-eq-5} therefore preserves the mean loss and scales the risk by $g_D=3D/[2(D+1)]$. The local ratio can approach $2+\sqrt2$, making Eq.~\eqref{app-product-channels-eq-21} sharp.

The optimal qubit weights $(p_0,p_+,p_1)=(3/10,2/5,3/10)$ give the local Haar coefficient $10/3$ (Appendix~\ref{app-probe-geometry-optimizing-all-sampling-weights}). The bound on $x_j$ has coefficients $5p_r$ after the same $3/2$ conversion, giving
\begin{equation}
\label{app-product-channels-eq-22}
x_j\le\frac32y_{j0}+2y_{j+}+\frac32y_{j1}.
\end{equation}

All three coefficients are at least one, as required by Eq.~\eqref{app-product-channels-eq-18}. Define the optimized global loss by $\epsilon_p=(3/10)Y_0+(2/5)Y_++(3/10)Y_1$. Applying the transfer inequality and multiplying by $D/(D+1)$ yields
\begin{equation}
\label{app-product-channels-eq-23}
R(\mathcal E)\le\frac{5D}{D+1}\epsilon_p(\mathcal E).
\end{equation}

Restricting noise to one site shows that any global weight choice has coefficient at least $g_D$ times the sharp local coefficient for those same weights. The local coefficient is at least $10/3$, so Eq.~\eqref{app-product-channels-eq-23} attains the best global coefficient $5D/(D+1)$ over all weights on this menu.

\subsection{Three-direction sample complexity and exact coefficient}
\label{app-product-channels-three-direction-sample-complexity-and-exact-coefficient}

Write $C_n=5D/(D+1)\le5$. The optimized test rejects with probability at least $\delta/C_n$ when the risk is at least $\delta$. The all-pass cap in Eq.~\eqref{main-cert-upper} is therefore of order $\delta^{-1}\log(1/\alpha)$ uniformly in $n$.

For the lower bound, apply $\mathcal D_q(X)=(1-q)X+q\Tr(X)I/2$ at one site and the identity elsewhere. The mixed component fails any pure-input survival test with probability $1/2$, so the local Haar risk is $q/2$ and the entanglement infidelity is $3q/4$. Choosing $q=4(D+1)\delta/(3D)\le1/4$ gives global risk $\delta$ by Eq.~\eqref{app-product-channels-eq-5}.

The survival loss is $b=q/2\le\delta$ on every global probe. Using $-\log(1-b)\le2\delta$ and the uniform confidence comparison in the fidelity bound~\eqref{main-fidelity-call-bound} gives the matching lower order. Thus
\begin{equation}
\label{app-product-channels-eq-24}
T_3^\star(\delta,\alpha)
=\Theta\!\left(\frac1\delta\log\frac1\alpha\right)
\end{equation}

uniformly in $n$ in the same small-error range.

For the exact coefficient, place the multi-Kraus adversary of Appendix~\ref{app-certification-exact-multi-kraus-adversary} at one site. The qubit gap $\mu_*=1/10$ gives $R_{\rm local}\ge t/3$ and $e_i\le t/10+t^2$ on all three rays. The single-site embedding preserves each probe loss and gives
\begin{equation}
\label{app-product-channels-eq-25}
R_{\rm global}=\frac{3D}{2(D+1)}R_{\rm local}.
\end{equation}

Choose $t=3\delta/g_D=2(D+1)\delta/D$, giving $R_{\rm global}\ge g_Dt/3=\delta$. Substitution into $t/10+t^2$ bounds every global probe loss by
\begin{equation}
\label{app-product-channels-explicit-coefficient-adversary}
 b_{D,\delta}=\frac{D+1}{5D}\delta
       +\frac{4(D+1)^2}{D^2}\delta^2,
\end{equation}
and for sufficiently small $\delta$, the lower denominator $-\log(1-b_{D,\delta})$ and the upper denominator $-\log(1-\delta/C_n)$ are both $\delta/C_n+O(\delta^2)$ at fixed $n$. The confidence ratio and ceiling estimates in Appendix~\ref{app-certification-exact-optimized-sampling-coefficient} then give, for fixed $n$ and ideal target projectors,
\begin{equation}
\label{app-product-channels-eq-26}
\lim_{\delta\downarrow0,\,\alpha\downarrow0}
\frac{\delta T_3^\star(\delta,\alpha)}{\log(1/\alpha)}
=\frac{5D}{D+1}.
\end{equation}

To reproduce Table~\ref{main-table-certification}, compute $x_\delta$ from Eq.~\eqref{app-product-channels-eq-12}. Substituting $t=(1-s^2)/2$ in Eq.~\eqref{app-product-channels-eq-11} gives $8x_\delta=5-4s-s^2$, so $s=\sqrt{9-8x_\delta}-2$. Compute $y_+=(1-s)^2/4$ and $b=1-(1-y_+)^n$, then round up the lower-bound ratio in Eq.~\eqref{app-product-channels-eq-16}. The three-probe column uses the all-pass ceiling at rejection probability $\delta/C_n$.

\subsection{A local physical realization of the hard channel}
\label{app-product-channels-a-local-physical-realization-of-the-hard-channel}

Use the local dilation in Eq.~\eqref{app-nonunitality-eq-19} at every site. The dilations can run in parallel, with a constant number of gates and one reset environment qubit per site and call. Tracing out each local environment gives the product of the reduced system outputs used in the fidelity bound.

\subsection{Physical resource scaling}
\label{app-product-channels-physical-resources-and-scope}

Each global call uses the $n$ local channels once and returns $n$ output qubits. Thus multiplying the two repetition laws by $n$ gives total local channel applications of
\begin{equation}
\label{app-product-channels-eq-29}
\Theta\!\left(n^2\delta^{-2}\log(1/\alpha)\right)
\quad\text{and}\quad
\Theta\!\left(n\delta^{-1}\log(1/\alpha)\right),
\end{equation}

respectively. The ideal product test is implemented by joint acceptance of the local projectors. With ideal primitives, preparation and measurement each take $O(n)$ one-qubit operations; gate synthesis and calibration enter the implementation budget. Each call of the SWAP comparison in Appendix~\ref{app-certification-trusted-pure-target-references} uses a fresh $n$-qubit target register. Doubling the leading call bound suffices, with both precision exponents preserved.

The product promise transfers the local linear bound to the whole register with three global probes. For arbitrary $D$-dimensional CPTP alternatives, linear robustness requires $N>D$ (Appendix~\ref{app-optimal-design}).

\clearpage
\twocolumngrid

\end{document}